\documentclass[hyper]{JHEP3}

\usepackage[dvips]{graphicx}
\usepackage{epsfig,multicol,amssymb,amsmath,cancel,subfigure}

\newcommand\fverb{\setbox\fverbbox=\hbox\bgroup\verb}
\newcommand\fverbdo{\egroup\medskip\noindent%
            \fbox{\unhbox\fverbbox}\ }
\newcommand\fverbit{\egroup\item[\fbox{\unhbox\fverbbox}]}
\newbox\fverbbox

\def\bea{\begin{eqnarray}}
\def\eea{\end{eqnarray}}
\def\beq{\begin{equation}}
\def\eeq{\end{equation}}

\title{Effective interactions of axion supermultiplet and thermal production of  axino dark matter}

\author{Kyu Jung Bae, Kiwoon Choi, Sang Hui Im  \\
Department of Physics, KAIST, Daejeon 305-701, Korea \\
E-mail : \email{kchoi@kaist.ac.kr}\\
\email{kyujung.bae@kaist.ac.kr}\\
\email{shim@muon.kaist.ac.kr}}

\abstract{We discuss the effective interactions of axion
supermultiplet, which might be important for analyzing the
cosmological aspect of supersymmetric axion model. Related to axino
cosmology, it is stressed that three seemingly similar but basically
different quantities, the Wilsonian axino-gluino-gluon coupling, the
1PI axino-gluino-gluon amplitude, and the PQ anomaly coefficient,
should be carefully distinguished from each other for correct
analysis of the thermal production of axinos in the early Universe.
It is then noticed that the 1PI axino-gluino-gluon amplitude at
energy scale $p$ in the range $M_\Phi < p < v_{PQ}$ is suppressed by
$M^2_\Phi/p^2$ in addition to the well-known suppression by
$p/16\pi^2 v_{PQ}$, where $M_\Phi$ is the mass of the heaviest
PQ-charged and gauge-charged matter supermultiplet  in the model,
which can be well below the PQ scale $v_{PQ}$.
 As a result, axino production at
temperature $T>M_\Phi$ is dominated by the production by matter
supermultiplet,  not
 by the production by
gauge supermultiplet.
Still the axino production rate is greatly reduced if $M_\Phi \ll
v_{PQ}$, which would make the subsequent cosmology significantly
altered. This would be most notable in the supersymmetric DFSZ model
in which $M_\Phi$ corresponds to the Higgsino mass which is around
the weak scale, however a similar reduction is possible in the KSVZ
model also. We evaluate the relic axino density for both the DFSZ
and KSVZ models while including the axino production in the
processes involving the heaviest PQ-charged  and gauge-charged
matter supermultiplet.}


\begin{document}

\section{\label{s1}Introduction}

Supersymmetric extension of the standard model provides perhaps the
most appealing solution to the gauge hierarchy problem \cite{susy}.
Furthermore it can easily accommodate the axion solution  to
 the strong CP
problem \cite{PQ, axion, strongcp}
which is another naturalness problem of the standard model.
Supersymmetric axion model \cite{axionsusy,kim_nilles} necessarily
contains the superpartners of axion, the axino and saxion, which can
have
 a variety of cosmological implications. In particular,
 axino can be a good candidate for cold dark matter \cite{cdm} in the Universe, depending upon the mechanism of
supersymmetry breaking and the presumed cosmological scenario
\cite{axino_dm,axino_cos}.

 One of the  key issues in axino cosmology  is the
thermal production of axino  by scattering or decay of particles in
thermal equilibrium  in the early Universe. Most of the previous
analysis of thermal axino production
\cite{axino_thermal_covi,Brandenburg,Strumia} is based on the
following form of local effective interaction between the axion
supermultiplet and the gauge supermutiplet: \bea \label{eff1} \int
d^2\theta \, \frac{1}{32\pi^2}
\frac{A}{v_{PQ}}W^{a\alpha}W^a_\alpha,\eea where $v_{PQ}$ is the
scale of spontaneous PQ breakdown,
and $A=(s+ia)/\sqrt{2}+\sqrt{2}\theta \tilde a +\theta^2 F^A$ is the axion
superfield which contains the axion $a$, the saxion $s$, and the
axino $\tilde a$ as its component fields.
If axinos were in thermal equilibrium, their relic density can be
determined in a straightforward manner, and the result does not
depend on the details of axino interactions. For the other case that
axinos were not in thermal equilibrium, one usually assumes that the
production is mostly due to the effective interaction (\ref{eff1}).
 Then a simple dimensional analysis
implies  that the axino production rate per unit volume scales as
$\Gamma_{\tilde a} \propto T^6/(16\pi^2 v_{PQ})^2$, which results in
a relic axino number to entropy ratio $Y_{\tilde a}\propto T_
R/(16\pi^2 v_{PQ})^2$, where the reheat temperature $T_R$ is assumed
to be lower than both $v_{PQ}$ and  the axino decoupling
temperature. If axino is a stable dark matter (DM), this result
leads to a severe upper limit on $m_{\tilde a}T_R$
\cite{axino_dm,axino_cos}. Even when axino is not stable, thermal
axino production in the early Universe can affect the later stage of
cosmological evolution in various ways, e.g. there can be a late
decay of axino into lighter particles in the minimal supersymmetric
standard model (MSSM) or into the lighter gravitino and axion
\cite{axino_decay1,cheung,axino_decay2,jasper}, which might affect
the relic DM density, or the Big-Bang nucleosynthesis, or the
structure formation.

In this paper, we discuss the effective interactions of axion
supermultiplet in a general context, and examine the implications
for cosmological axino production. It is stressed that three
seemingly similar but basically  different  quantities,  the
Wilsonian axino-gaugino-gauge boson coupling, the 1PI
axino-gaugino-gauge boson amplitude, and the PQ anomaly coefficient,
should be carefully distinguished from each other for  correct
analysis of the cosmological axino production. At any given scale,
the Wilsonian coupling is local by construction, but it depends
severely on the field basis adopted to define the effective theory.
On the other hand, the 1PI amplitude is an observable quantity, and
therefore is basis-independent, while it  generically contains
non-local piece and has different value at different energy scale of
external particles. The PQ anomaly coefficient is an intrinsic
property of the PQ symmetry of the model, which is basis-independent
and has a common value at all scales as a consequence of the anomaly
matching condition.

At any temperature below $v_{PQ}$, axino production by gauge
supermultiplet  is determined by the 1PI axino-gaugino-gauge boson
amplitude at the corresponding energy scale.
 Our key observation is that  the 1PI axino-gaugino-gauge boson
amplitude at energy scale $p$ in the range
 $M_\Phi < p < v_{PQ}$  does not scale like the
Wilsonian  effective interaction (\ref{eff1}),
but is suppressed further by the factor $M_\Phi^2/p^2$, where
$M_\Phi$ is the mass of
 the heaviest
PQ-charged and gauge-charged matter field  in the model, which can
be well below $v_{PQ}$ in general. This is a simple consequence of
that the axion supermultiplet is decoupled from gauge
supermultiplets (and from charged matter supermultiplets also) in
the limit $M_\Phi\rightarrow 0$, which is not manifest if one
considers the effective interaction (\ref{eff1}) alone, but
recovered in the full analysis taking into account all interactions
of the axion supermultiplet together.
As a result, if $M_\Phi\ll v_{PQ}$, which is possible in most cases,
axino production by gauge
 supermultiplets is greatly reduced.
  This would be most dramatic in
the supersymmetric Dine-Fishler-Srednicki-Zhitnitsky (DFSZ) model
\cite{dfsz} as the model does not contain any exotic PQ-charged and
gauge-charged matter field other than the matter and Higgs fields in
the MSSM, and therefore $M_\Phi$ is around the weak scale.
 A similar suppression is possible in the
 supersymmetric Kim-Shifman-Vainshtein-Zakharov (KSVZ) axion model \cite{ksvz} also, in which
  $M_\Phi$ corresponds to the mass of
  an exotic PQ-charged quark multiplet, which in principle can take   any value
between the PQ scale and the weak scale.

 An immediate consequence
of the above observation is that axino production in the temperature
range $M_\Phi <T <v_{PQ}$ is dominated {\it not} by the production
by  gauge supermultiplets, {\it but} by the production by the
heaviest PQ-charged and gauge-charged matter multiplet $\Phi$. If
$M_\Phi \ll v_{PQ}$,  axino production rate at $T>M_\Phi$ is greatly
 reduced compared to the previous result based on the effective interaction (\ref{eff1}) alone.
 This can significantly alter the cosmological aspect of the model. For instance, a
high reheat temperature which has been considered to produce  too
much axino dark matter can be allowed  if $M_\Phi\ll v_{PQ}$. Also,
our result can relieve the upper bound on the reheat temperature,
which has been  obtained in \cite{cheung} by considering the
gravitino and axino productions together, and might have implication
for the cosmological bounds on axion supermultiplet in the presence
of small $R$-parity breaking, which have been discussed in
\cite{jasper}. Since the interaction rate between the axino and the
thermal bath of gauge-charged particles is reduced in the limit
$M_\Phi \ll T$, the axino decoupling temperature can take a higher
value than the previous estimate based on (\ref{eff1}). Similar
observations apply to the axion and saxion cosmology also, and our
result might affect some of the results of the recent analysis of
thermal axion production \cite{axion_thermal}.

The organization of this paper is as follows. In the next section,
we discuss in a general context the effective interactions of axion
supermultiplet which will be crucial  for later discussion of
cosmological axino production. In section 3, we provide an analysis
of thermal axino production for both the KSVZ and DFSZ models, and
determine the relic axino density including the contribution from
the processes involving the heaviest PQ-charged  and gauge-charged
matter multiplet at $T>M_\Phi$.  Section 4 is the conclusion.

\section{Effective interactions  of axion supermultiplet}

In this section, we discuss the generic feature of the effective
interactions of axion supermultiplet \cite{axion_eff}, which will be
relevant for our later discussion of cosmological axino production.
For the purpose of illustration,
we first consider a simple specific model, a supersymmetric
extension of the KSVZ axion model \cite{ksvz}, and later generalize
the discussion to generic supersymmetric axion models.

  At the fundamental scale $M_*$, which is presumed to be
of the order of the reduced Planck scale $M_{Pl}\simeq 2.4\times
10^{18}$ GeV,
the supersymmetric part of our model is described by
 the following
high scale lagrangian \bea \label{lag0} {\cal L} (M_*) =\int d^4
\theta\, K +\left[ \int d^2\theta \, \left(\frac{1}{4} f_a
W^{a\alpha}W^a_{\alpha} + W(\Phi_I)\right)+{\rm h.c.}\,\right],\eea
where $W^a_\alpha$ are the gluon superfields, and the K\"ahler
potential $K$, the holomorphic gauge kinetic function $f_a$, and
the superpotential $W$ are given by \bea \label{uv1} K &=& \sum_I
\Phi_I^\dagger \Phi_I, \qquad f_a \, = \,
\frac{1}{\hat g_s^2(M_*)},\nonumber \\
W&=& h Z\left(XY-\frac{v_{PQ}^2}{2}\right)+\lambda X QQ^c, \eea
where $\hat g_s^2(M_*)$ is the holomorphic QCD coupling at $M_*$,
 and $\Phi_I$ stand for the chiral matter superfields in the model,
 $\{\Phi_I\}=\{Z,X,Y,Q,Q^c\}$.
 Among these matter fields, $Z, X$ and $Y$ are gauge singlet,
while $Q+Q^c$ are colored quark multiplet. Here we distinguish the
holomorphic QCD coupling $\hat g_s$ from the physical QCD coupling
$g_s$ for later discussion.
Also we ignore non-renormalizable operators
suppressed by $1/M_*$, as well as the similarly suppressed
supergravity effects, under the assumption that the PQ scale
$v_{PQ}$ is far below $M_*$, so that any effect suppressed by
$v_{PQ}/M_*$ can be safely ignored.


The model is invariant under the global PQ symmetry\footnote{In
addition to the PQ symmetry, we can introduce an approximate
$U(1)_R$ symmetry under which $\theta^\alpha \rightarrow e^{i\beta}
\theta^\alpha$ and $Z\rightarrow e^{2i\beta}Z$ to justify the form
of the superpotential. In the present form, the model is invariant
under another $U(1)$  symmetry which makes the massive Dirac quark
$Q+Q^c$ stable. However it is straightforward to break this $U(1)$,
while keeping the PQ symmetry unbroken, when the model is  combined
with the MSSM, making $Q+Q^c$ decay fast enough.}\bea \label{pq0}
U(1)_{PQ}: \quad \Phi_I \rightarrow e^{ix_I\alpha} \Phi_I,\eea where
the PQ charge $x_I$ are given by \bea
\big(x_Z,x_X,x_Y,x_Q+x_{Q^c}\big) = \big(0, -1, 1, 1\big).\eea
This PQ symmetry is explicitly broken by the QCD anomaly 
as\bea
\partial_\mu J^\mu_{PQ}=\frac{g^2}{16\pi^2}C_{PQ}F^{a\mu\nu}\widetilde{F}^a_{\mu\nu},\eea
where $J^\mu_{PQ}$ is the Noether current for the symmetry
transformation (\ref{pq0}) defined in a gauge invariant
regularization scheme, and then the anomaly coefficient is given by
\bea \label{pqano} C_{PQ}=2\sum_I x_I {\rm Tr}(T_c^2(\Phi_I)) =2
(x_Q+x_{Q^c}){\rm Tr}(T_c^2(Q))= N_Q,\eea where $T_c(\Phi_I)$ is the
generator of the $SU(3)_c$ transformation of $\Phi_I$, and $N_Q$ is
the number of $Q+Q^c$.
The QCD instantons 
then generate an axion potential which has a minimum at the axion
vacuum expectation value (VEV) which cancels the QCD vacuum angle, so the strong CP problem is
solved \cite{strongcp}. In our convention, the PQ anomaly
coefficient $C_{PQ}$ corresponds to the number of the discrete
degenerate vacua of the axion potential, which can have a nontrivial
cosmological implication \cite{axiondomain}. In the following, we
will assume $N_Q=1$ for simplicity.

In the limit to ignore SUSY breaking effects, the model has a
supersymmetric ground state \bea \label{f-flat} \langle XY \rangle
=\frac{v_{PQ}^2}{2},\quad \langle Z\rangle =0.\eea  Including SUSY
breaking effects, the VEV of the ratio
$X/Y$ can be fixed and a small but nonzero VEV of $Z$ can be induced
also. In fact, SUSY breaking part of the model is not essential for
our discussion of axino effective interactions, although it is
crucial for the determination of the axino and saxion masses
\cite{axino_mass_old,axino_mass_new}. We thus simply assume that
soft SUSY breaking terms in the model fix the VEVs  as \bea \langle
X \rangle \sim \langle Y\rangle \eea with appropriate
(model-dependent, but light) axino and saxion masses, for which the
axion scale is given by $v_{PQ}$. Also, to illustrate our main
point, we further assume that $h={\cal O}(1)$, but $\lambda\ll 1$,
so the model involves two widely separated mass scales, the axion
scale $v_{PQ}$ and the quark mass $M_Q $ which can be far below
$v_{PQ}$: \bea M_Q\equiv \frac{\lambda v_{PQ}}{\sqrt{2}}\, \ll \,
v_{PQ}.\eea

All physical consequences   of the model can be determined in
principle by the high scale lagrangian (\ref{lag0}).
However, for low energy dynamics at scales below $v_{PQ}$, it is
convenient to construct an effective lagrangian at a cutoff scale
$\Lambda$ {\it just below} $v_{PQ}$, which can be obtained by
integrating out the massive fields heavier than $\Lambda$ as well as
the high momentum modes of light fields at scales above $\Lambda$.
In our case,
 two chiral superfields among $X$, $Y$ and $Z$ get a mass of ${\cal
O}(v_{PQ})$,
while the remained combination provides the axion supermultiplet.
To integrate out the heavy fields, one can parameterize them as \bea
X=\frac{1}{\sqrt{2}}\left(v_{PQ}+\rho_1\right)e^{-A/v_{PQ}}, \quad Y
=\frac{1}{\sqrt{2}}\left(v_{PQ}+\rho_1\right)e^{A/v_{PQ}}, \quad
Z=\rho_2, \eea for which the  K\"ahler potential and superpotential
 at the UV scale $M_*$ are given by
 \bea \label{uv2} K &=& |v_{PQ}+\rho_1|^2\cosh\left(\frac{A+A^\dagger}{v_{PQ}}\right)+\rho_2^\dagger\rho_2+Q^\dagger
Q + Q^{c\dagger} Q^c,\nonumber \\
W&=&h v_{PQ}\rho_1\rho_2 + \frac{1}{2}
h\rho_2\rho_1^2+\left(M_Q+\sqrt{2}\frac{M_Q}{v_{PQ}}\rho_1\right)e^{-A/v_{PQ}}QQ^c,
\eea where $\rho_i$ ($i=1,2$) are
massive chiral matter fields, and \bea A =
\frac{1}{\sqrt{2}}\left(s+ia\right) + \sqrt{2}\theta \tilde a + \theta^2
F^A\eea
 is the axion
superfield  which contains the axion $a$,  the saxion $s$ and the
axino $\tilde{a}$ as its component fields.

After integrating out the heavy $\rho_i$ and the high momentum modes
of light fields, the resulting Wilsonian effective lagrangian can be
chosen to  take the following form \bea \label{lag1} {\cal L}_{\rm
eff} (\Lambda) &=& \int d^4 \theta\,
\left[\,K_A(A+A^\dagger)+Z_Q(A+A^\dagger)Q^\dagger
Q+Z_{Q^c}(A+A^\dagger)Q^{c\dagger}Q^c\,\right]\nonumber \\
&& +\,\left[ \int d^2\theta \, \left(\frac{1}{4} f^{\rm eff}_a(A)
W^{a\alpha}W^a_{\alpha} + W_{\rm eff}\right)+{\rm h.c.}\,\right],\eea
 where \bea K_A
&=& \frac{1}{2}(A+A^\dagger)^2+{\cal
O}\left(\frac{(A+A^\dagger)^3}{v_{PQ}}\right)\nonumber \\
 f^{\rm eff}_a & = &
\frac{1}{\hat g_s^2(\Lambda)}-\frac{C_W}{8\pi^2}\frac{A}{v_{PQ}} ,
\nonumber \\
W_{\rm eff}&=& M_Q e^{-(\tilde x_Q+\tilde x_{Q^c})A/v_{PQ}} QQ^c.
\eea  With the above form of effective lagrangian, the PQ symmetry
is realized as \bea \label{pqeff} U(1)_{PQ}:\quad A\rightarrow A
+i\alpha v_{PQ}, \quad \Phi \rightarrow e^{i \tilde
x_\Phi\alpha}\Phi \quad (\Phi=Q,Q^c),\eea and  the PQ anomaly in the
effective theory is given by \bea C_{PQ}= C_W + \tilde x_Q + \tilde
x_{Q^c},\eea which should be same as the PQ anomaly (\ref{pqano}) in
the underlying UV theory. For generic form of effective lagrangian
consistent with the PQ symmetry (\ref{pqeff}), there can be three
types of axino effective interactions at the order of $1/v_{PQ}$:
\bea
 \label{e11}
\Delta_1 {\cal L}(\Lambda) &=& -\int d^2\theta
\,\frac{C_W}{32\pi^2}\frac{A}{v_{PQ}}W^{a\alpha}W^a_\alpha+{\rm
h.c.},
 \\
 \label{e22}
\Delta_2 {\cal L}(\Lambda)&=&\int d^4\theta \,
\frac{(A+A^\dagger)}{v_{PQ}} \left( \tilde y_Q Q^\dagger Q +\tilde
y_{Q^c}Q^{c\dagger}Q^c\right),
 \\
\label{e33}\Delta_3 {\cal L}(\Lambda)&=&-\int d^2\theta \,(\tilde
x_Q+\tilde x_Q^c)M_Q\frac{A}{v_{PQ}}QQ^c +{\rm h.c.} ,\eea where \bea
\tilde y_{\Phi} &\equiv & v_{PQ}\left.\frac{\partial \ln
Z_{\Phi}}{\partial A}\right|_{A=0} \quad (\Phi=Q,Q^c).\eea  In our
particular case, one easily finds \bea C_W=0, \quad \tilde
x_{\Phi}=x_{\Phi},\quad \tilde y_{\Phi}(\Lambda) ={\cal
O}\left(\frac{1}{16\pi^2} \frac{M_Q^2}{v_{PQ}^2}\ln
\left(\frac{M_*}{\Lambda}\right)\right), \eea where the small
$\tilde y_\Phi$ is induced by the radiative corrections involving
the Yukawa coupling $\lambda = \sqrt{2}M_Q/v_{PQ}$ at UV scales
above $\Lambda$. This result shows that the axion supermultiplet is
decoupled from the gauge multiplet and charged-matter multiplet in
the limit $M_Q\rightarrow 0$, which is manifest in the UV theory
(\ref{uv2}). Note  that the PQ symmetry splits into two global
$U(1)$ symmetries in the limit $M_Q\rightarrow 0$, the axial $U(1)$
symmetry of $Q, Q^c$ and the anomaly free $U(1)$ not involving the
transformation of $Q, Q^c$, and the axion becomes a massless
Goldstone boson of the anomaly-free $U(1)$ part in the limit
$M_Q\rightarrow 0$.


In fact, one can consider a different description of the same
effective theory, for instance a scheme  using  different form of PQ
symmetry under which all light fields at $\Lambda$ are invariant
except the axion superfield. In our case, such description
can be achieved by the $A$-dependent field redefinition $ \Phi
\rightarrow e^{x_\Phi A/v_{PQ}} \Phi$, making all matter fields
except $A$ invariant under the PQ symmetry. Such form of PQ symmetry
can be convenient in certain respects since $\tilde x_\Phi=0$
($\Phi=Q,Q^c$), so the PQ anomaly originates entirely from the
variation of the local effective interaction (\ref{e11}), which
results in $C_{PQ}=C_W$. However such field basis is not convenient
for a discussion of physics at energy scales above $M_Q$ since the
decoupling of the axion supermultiplet from the gauge and
charged-matter multiplets  in the limit $M_Q\rightarrow 0$ is not
manifest, but is achieved by the cancellation between the
contributions from $C_W=N_Q$ and $\tilde y_Q+\tilde y_{Q^c}=1$.

To get more insights, let us consider a more general description
associated with the field redefinition \bea \label{redef1}
 \Phi \rightarrow
e^{cx_\Phi A/v_{PQ}} \Phi \quad (\Phi=Q,Q^c), \eea
 where $c$ is an arbitrary real constant parameterizing the field basis.
 After this field redefinition,  the PQ symmetry is realized
as \bea \label{pq1} U(1)_{PQ}:\quad A\rightarrow A +i\alpha v_{PQ},
\quad \Phi \rightarrow e^{i(1-c)x_\Phi \alpha} \Phi, \eea and there
appears an axion-dependent local counter term in the effective
lagrangian \bea \label{wz} -\int
d^2\theta\,\frac{c(x_Q+x_{Q^c})}{32\pi^2}
\frac{A}{v_{PQ}}W^{a\alpha}W^a_\alpha+{\rm h.c.}, \eea  which is due
to the Konishi anomaly \cite{Konishi} for the  field redefinition (\ref{redef1}).
Note that this Konishi anomaly term is required to match the PQ
anomaly of (\ref{pq1}) with the PQ anomaly (\ref{pqano}).
 Including the Konishi anomaly, Wilsonian effective
 couplings of the axion superfield
 in the redefined field basis are given by \bea C_W  &=&  c(x_Q+x_{Q^c}),
\quad \tilde x_{\Phi}\,=\, (1-c) x_{\Phi},\nonumber \\
 \tilde
y_{\Phi}(\Lambda)&=& cx_{\Phi}+{\cal O}\left(\frac{1}{16\pi^2}
\frac{M_Q^2}{v_{PQ}^2}\ln
\left(\frac{M_*}{\Lambda}\right)\right).\eea
Note that the PQ anomaly in the redefined field basis  is determined
by two contributions: \bea C_{PQ}=
c(x_Q+x_{Q^c})+(1-c)(x_Q+x_{Q^c}),\eea where the first piece is from
the variation of the local effective interaction (\ref{wz}),
while the second piece is from the axial
anomaly of the PQ transformation of $Q,Q^c$ in (\ref{pq1}).
Obviously each contribution depends on the field basis parameter
$c$, but their sum is independent of $c$ as it should be.

Now, if one ignores the part of $\tilde y_\Phi$ suppressed by
$M_\Phi^2/v_{PQ}^2$,   Wilsonian effective interactions of the axion
supermultiplet   are given by
 \bea
 \label{e1}
\Delta_1 {\cal L}(\Lambda) &=& -\int d^2\theta
\,\frac{c(x_Q+x_{Q^c})}{32\pi^2}\frac{A}{v_{PQ}}W^{a\alpha}W^a_\alpha,+{\rm
h.c.},
 \\
 \label{e2}
\Delta_2 {\cal L}(\Lambda)&=&\int d^4\theta \,
\frac{(A+A^\dagger)}{v_{PQ}} \left(cx_Q Q^\dagger Q +
cx_{Q^c}Q^{c\dagger}Q^c\right)
 \\
\label{e3}\Delta_3 {\cal L}(\Lambda)&=&-\int d^2\theta
\,(1-c)(x_Q+x_Q^c)M_Q\frac{A}{v_{PQ}}QQ^c +{\rm h.c.}.\eea With
these, one can compute the  axino production rate at any temperature
$T<\Lambda$, and the result should be independent of the field basis
parameter $c$ since all physical quantities should be
basis-independent.  However, most of the previous studies use only
the effective interaction of the form (\ref{e1}), which can lead to
a misleading result as we will see below.
In fact, considering  only the effective interaction (\ref{e1})
leads to a highly overestimated result for the axino production at
energy scale $p\gg M_Q$. At such high energy scale, one can take a
limit $M_Q/p\rightarrow 0$ in which the axion supermultiplet is
decoupled from the gauge multiplets and charged-matter multiplets.
Although such decoupling is not manifest in the effective
interactions $(\ref{e1})-(\ref{e3})$ with $c\neq 0$, it is manifest
in the field basis with $c=0$,  and also in the underlying UV theory
(\ref{uv2}). This means that axino production by gauge
supermultiplets at $p
>M_Q$ should be suppressed by some powers of $M_Q/p$ if one
takes into account the interactions $(\ref{e1})-(\ref{e3})$
altogether to get a correct $c$-independent result.
  On the other hand, the interaction (\ref{e1}) by itself does
not involve any suppression by $M_Q/p$, and therefore the analysis
using (\ref{e1}) alone gives a highly overestimated result in the
limit $p\gg M_Q$.

More explicitly,
 the effective interaction (\ref{e1}) gives
 the Wilsonian axino-gluino-gluon amplitude  \bea \label{wil} {\cal A}_W(k,q,p)
= -\frac{g^2c(x_Q+x_{Q^c})}{16\pi^2 \sqrt{2}v_{PQ}}\delta^4(k+q+p)
\bar{u}(k)\sigma_{\mu\nu} \gamma_5 v(q)\epsilon^\mu p^\nu, \eea
where $p^\mu$ and $\epsilon^\mu$ are the gluon momentum and
polarization, and
$u(k)$ and $v(q)$ are the 4-component Majorana spinor wavefunction
of the axino and gluino, respectively.
\begin{figure}
\begin{center}
\subfigure[\label{proc_A_1}]{\includegraphics[width=3.5cm]{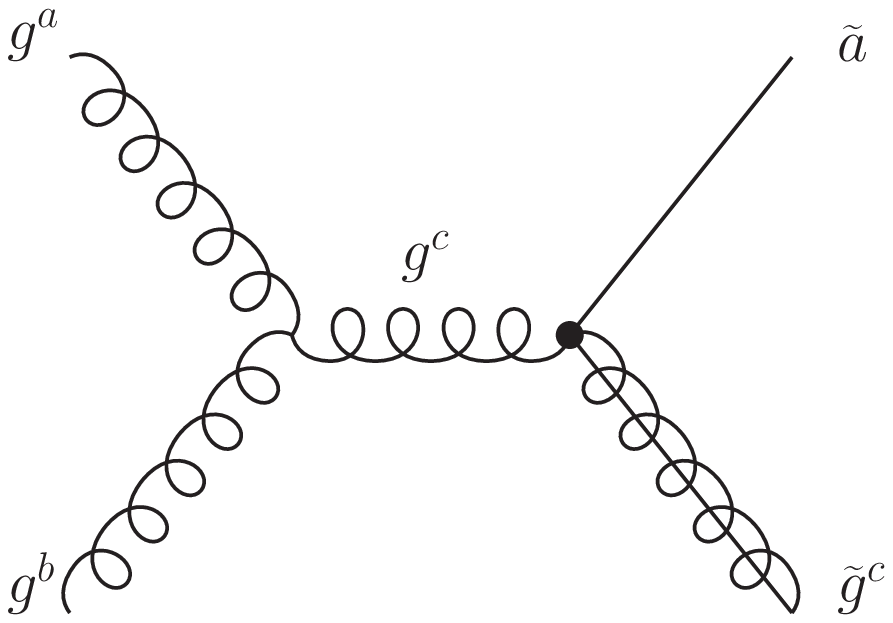}}
\quad
\subfigure[\label{proc_A_2}]{\includegraphics[width=3.5cm]{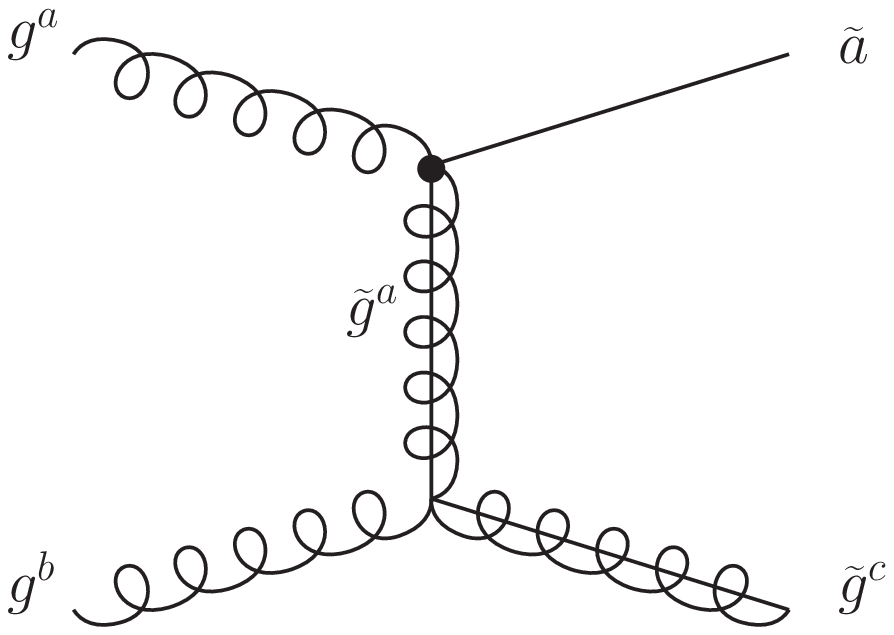}}
\quad
\subfigure[\label{proc_A_3}]{\includegraphics[width=3.5cm]{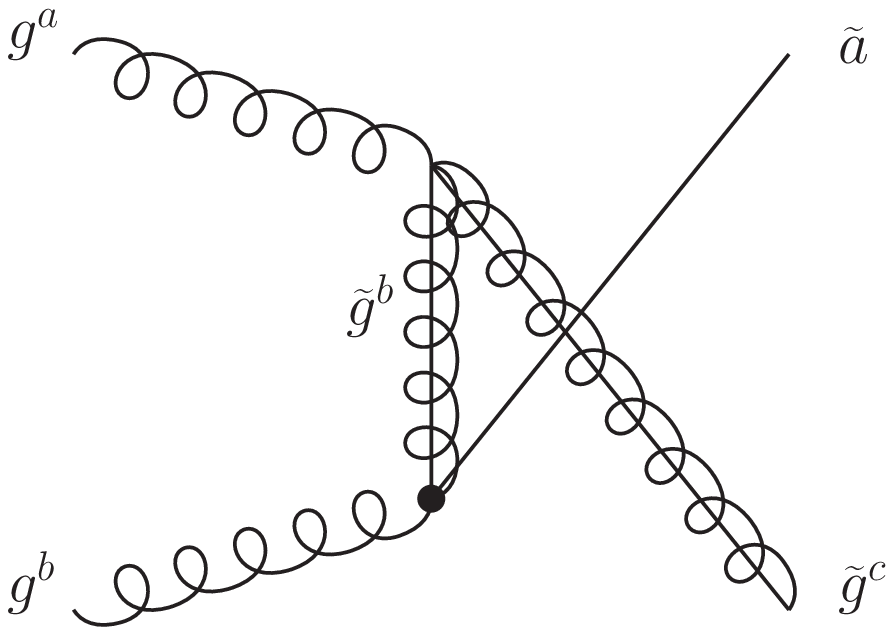}}
\caption{Diagrams for the process $g+g\rightarrow \tilde{g} +
\tilde{a}$. \label{proc_A_g}}
\end{center}
\end{figure}
    If one uses this amplitude alone to compute the axino
production rate in the process $\mbox{gluon} + \mbox{gluon}
\rightarrow \mbox{gluino}+\mbox{axino}$ (see Fig. \ref{proc_A_g}), a simple
dimensional analysis tells that
 the rate (per unit volume) is
given by \bea \Gamma(gg\rightarrow \tilde g{\tilde a})\, =\,
c^2(x_Q+x_{Q^c})^2\,\frac{\xi g_s^6T^6}{(16\pi^2 v_{PQ})^2},\eea
where $\xi$ is a dimensionless coefficient which is independent of
$c$. However this can not be the correct answer as it depends on the
field-basis parameter $c$, and there should be additional
contribution which removes the $c$-dependence of the result.
\begin{figure}
\begin{center}
\subfigure[\label{gga_1}]{\includegraphics[width=3.4cm]{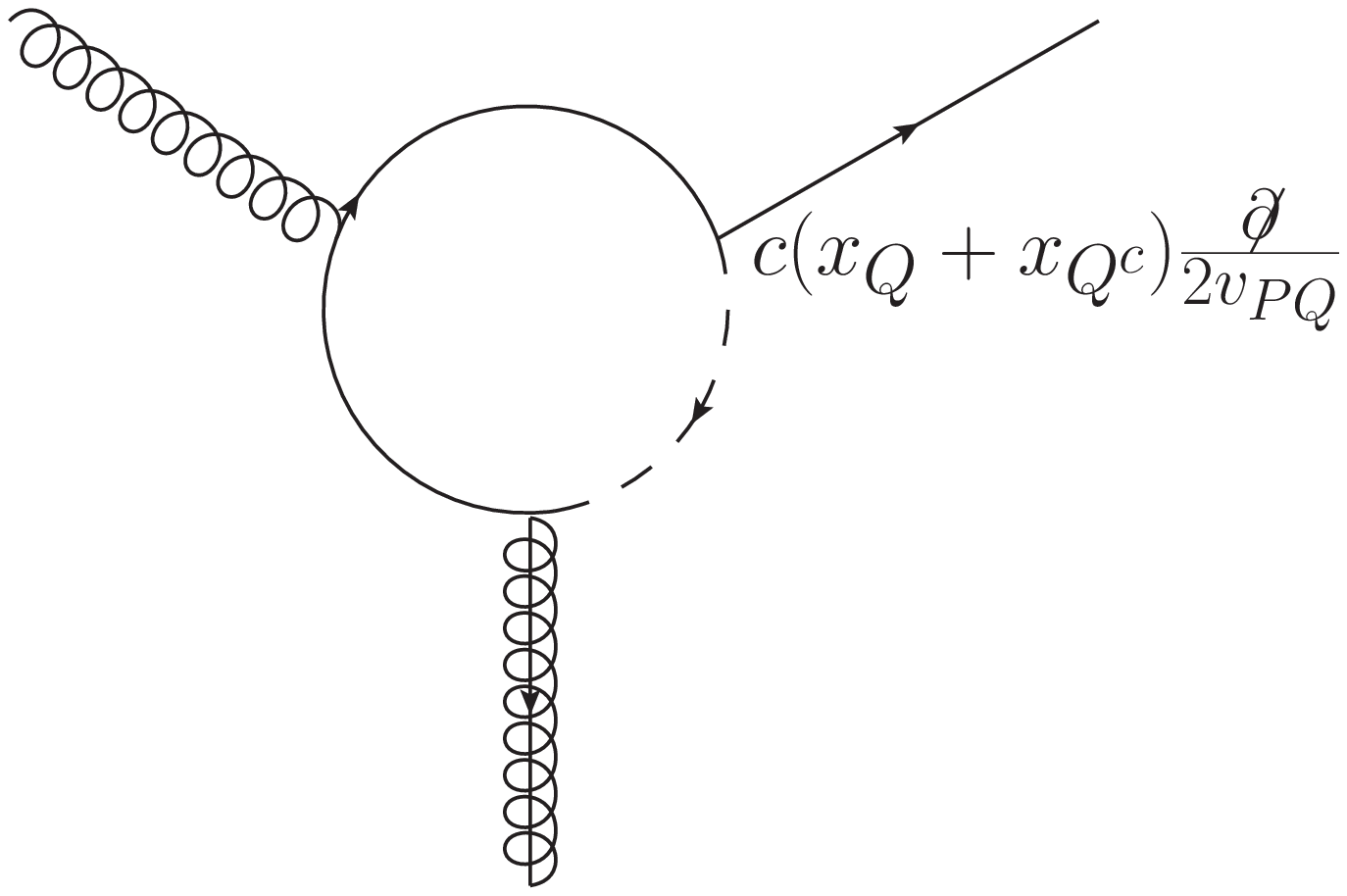}} \quad
\subfigure[\label{gga_2}]{\includegraphics[width=3.4cm]{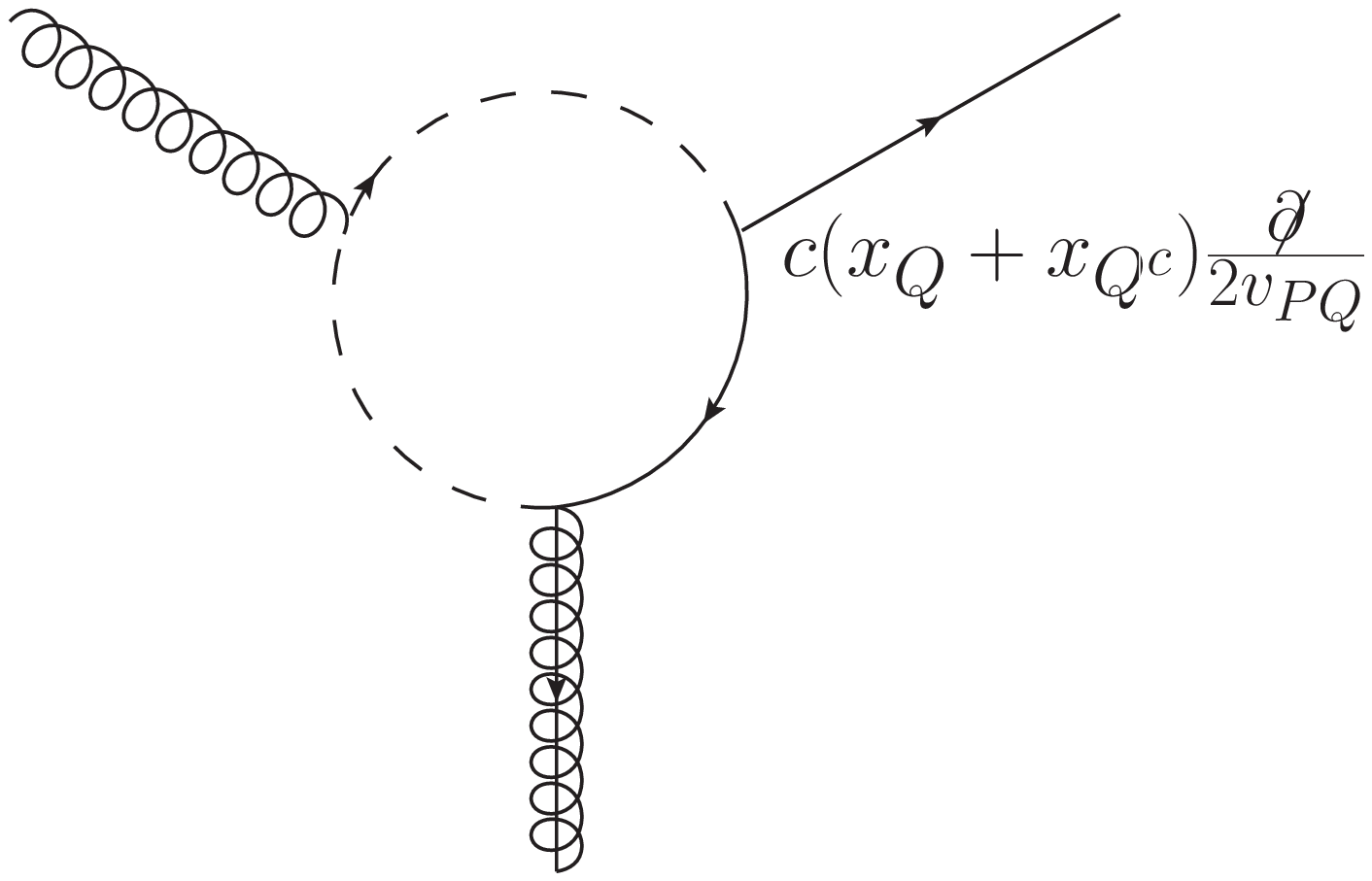}} \quad
\subfigure[\label{gga_3}]{\includegraphics[width=3.4cm]{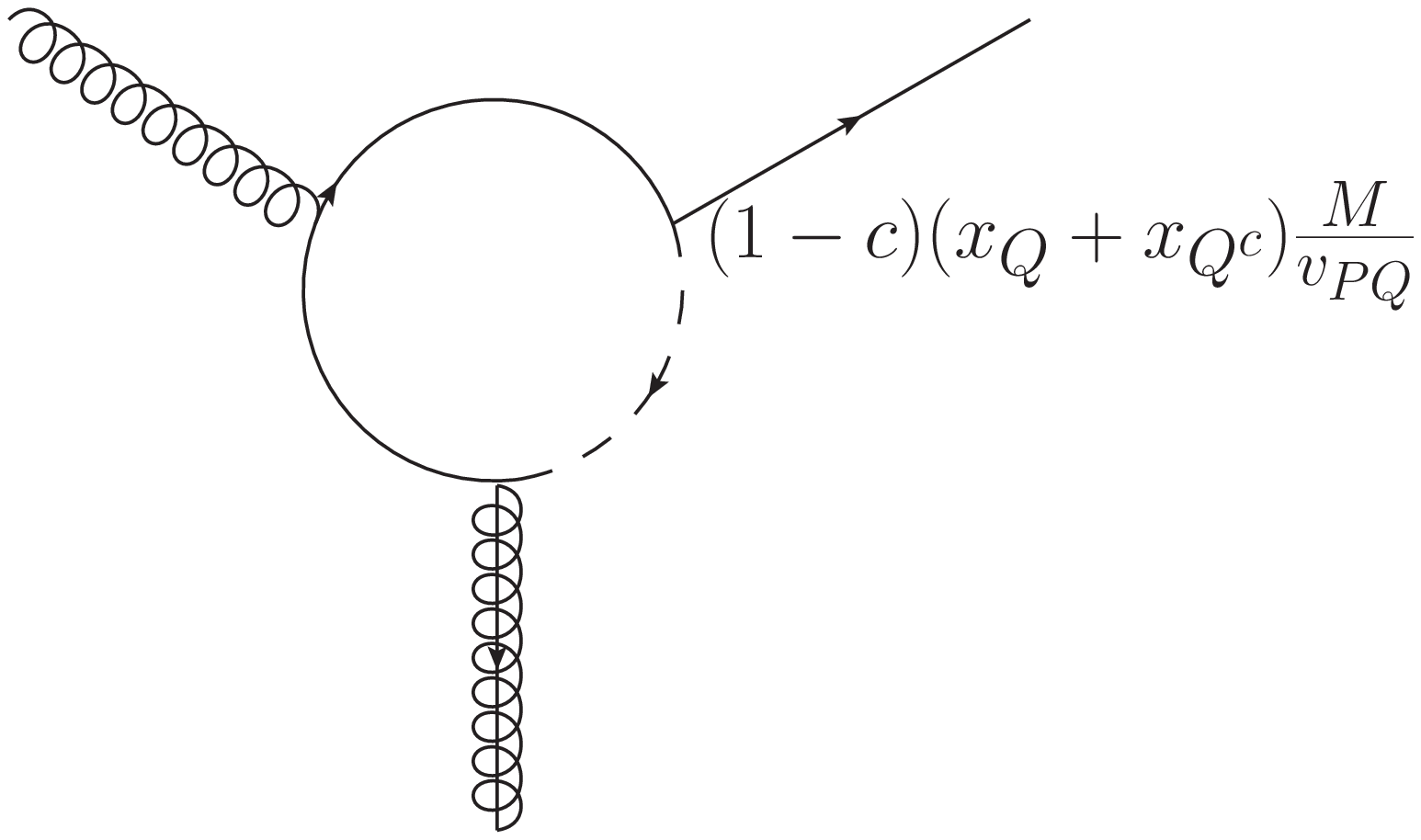}} \quad
\subfigure[\label{gga_4}]{\includegraphics[width=3.4cm]{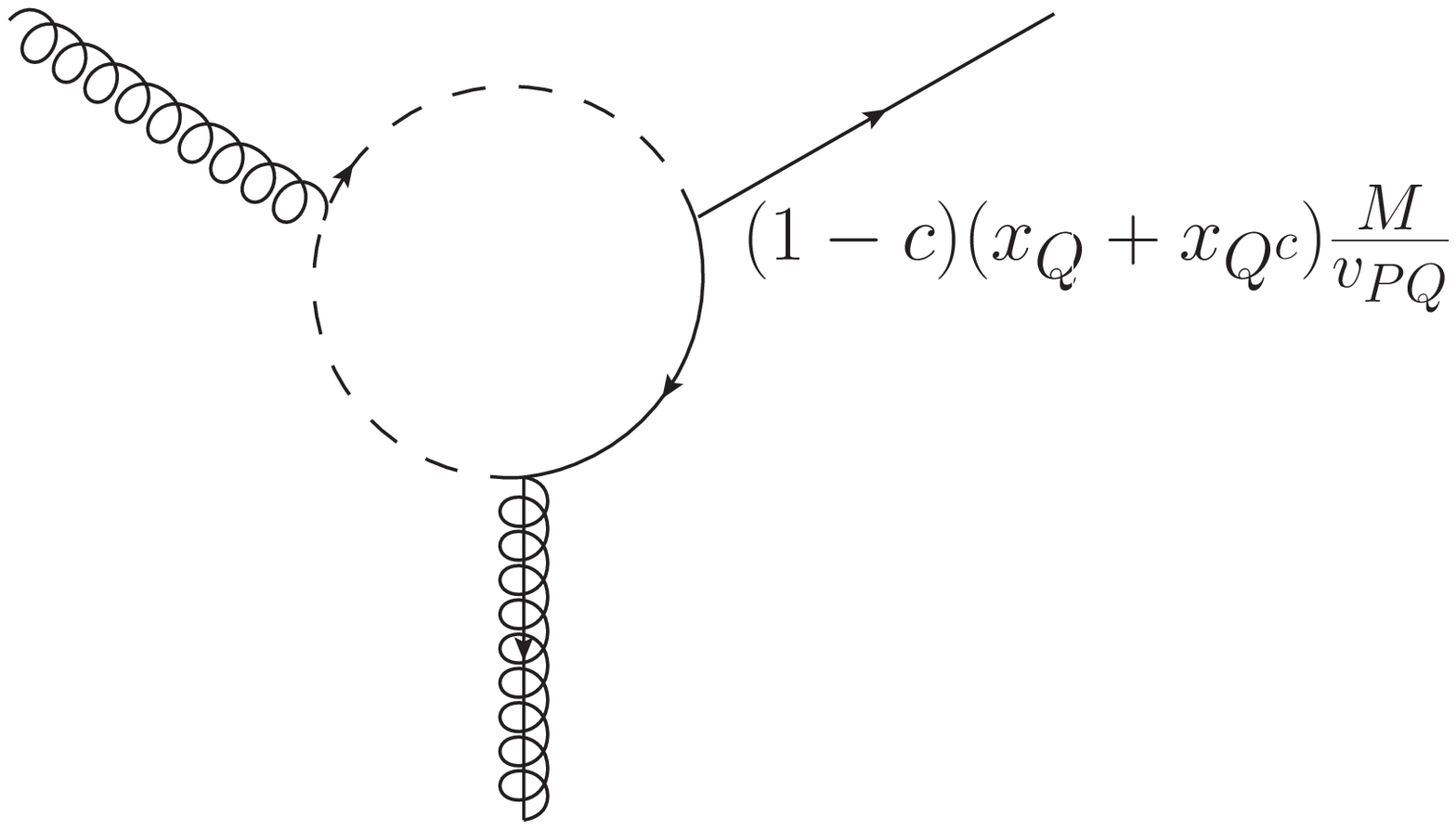}}
\caption{Contributions to the 1PI axino-gluino-gluon  amplitudes
from the loops of $Q, Q^c$. \label{gga}}
\end{center}
\end{figure}
Indeed, including the contributions from the  axino-gluon-gluino
amplitude due to the loops of $Q,Q^c$ (see Fig. \ref{gga}), which
involve the $c$-dependent axino-matter couplings in (\ref{e2}) and
(\ref{e3}),
we find that the $c$-dependence in the production rate disappears
as required, and the final result takes the form \bea \Gamma
(gg\rightarrow \tilde g{\tilde a}) \, = \, \gamma^2 \frac{\xi
g_s^6T^6}{(16\pi^2v_{PQ})^2},
\eea where \bea \gamma &=& {\cal
O}\left(\frac{M^2_Q}{T^2}\ln^2\left(\frac{T}{M_Q}\right)\right)
\quad \mbox{for} \quad T>M_Q,
\nonumber \\
\gamma &=& (x_Q+x_Q^c)+{\cal O}\left(\frac{T^2}{M^2_Q}\right) \quad
\mbox{for} \quad T<M_Q, \eea which shows that the production rate is
indeed suppressed by $M_Q^2/T^2$ at $T>M_Q$. This result can be
obtained by replacing the axino-gluino-gluon vertex in Fig.
\ref{proc_A_g} with the one-particle-irreducible (1PI)
axino-gluino-gluon amplitude \bea {\cal A}_{1PI}(k,q,p) =
-\frac{g^2\widetilde C_{1PI}(k,q,p)}{16\pi^2
\sqrt{2}v_{PQ}}\delta^4(k+q+p) \bar{u}(k)\sigma_{\mu\nu} \gamma_5
v(q)\epsilon^\mu p^\nu \eea
 which has  the following limiting
behavior (see the appendix A) when the axino and gluino are on
mass-shell: \bea \label{1pi1} \widetilde C_{1PI}(k^2=q^2=0; \, |p^2|\gg
M_Q^2) &=& (x_Q+x_Q^c)
\frac{M^2_Q}{|p^2|}\left[\ln^2\left(\frac{|p^2|}{M_Q^2}\right)
+{\cal
O}\left(\frac{M_Q^2}{|p^2|}\ln\left(\frac{|p^2|}{M_Q^2}\right)\right)\,\right], \nonumber\\
\label{1pi2} \widetilde C_{1PI}(k^2=q^2=0;\, |p^2|\ll M_Q^2) &=&
x_Q+x_Q^c +{\cal O}\left(\frac{p^2}{M_Q^2}\right).
 \eea

 Generically the 1PI amplitude
due to light particle loops includes non-local and non-analytic
piece in the limit $p^2\gg M_{\rm light}^2$, while the contribution
from heavy particle loop allows a local expansion in the limit
$p^2\ll M_{\rm heavy}^2$. For the 1PI axino-gaugino-gauge boson
amplitude, it takes the following form (see the appendix A) when the
axino and gaugino (or gauge boson) are on mass-shell: \bea \widetilde
C_{1PI}=C_{1PI}+{\cal O}\left(\frac{M^2_{\rm
light}}{p^2}\ln^2\left(\frac{p^2}{M_{\rm
light}^2}\right)\right)+{\cal O}\left(\frac{p^2}{M^2_{\rm
heavy}}\right), \eea where  $C_{1PI}$ can have a logarithmic
$p$-dependence due to higher loop effects.  The constant (or
logarithmic)  $C_{1PI}$ can be encoded in the PQ-invariant real 1PI
gauge kinetic function ${\cal F}_a$ which defines the 1PI gauge
kinetic term \cite{nsvz,arkani}: \bea \int d^4 \theta \,
\frac{1}{16}{\cal F}_a(\ln p^2,
A+A^\dagger)W^{a\alpha}\frac{D^2}{p^2} W^a_\alpha, \eea and then
$C_{1PI}^a$ is determined as \bea C_{1PI} = -8\pi^2
v_{PQ}\left.\frac{\partial {\cal F}_a}{\partial A}
\right|_{A=0}.\eea This 1PI operator
 should be distinguished from the Wilsonian gauge kinetic term
\bea  \label{wz1} \int d^2\theta\, \frac{1}{4}f_a^{\rm
eff}(\ln\Lambda,A)W^{a\alpha}W^a_\alpha +{\rm h.c.}\eea which
determines $C_W$ as \bea C_W = -8\pi^2 v_{PQ}\left.\frac{\partial
f^{\rm eff}_a}{\partial A} \right|_{A=0}.\eea
 As we have
noticed,  $C_W$ is a basis-dependent Wilsonian coupling  which
changes under an  $A$-dependent holomorphic redefinition of charged
matter fields, while $C_{1PI}$ is an observable amplitude invariant
under the redefinition of matter fields.

In the above, we find in the context of a simple supersymmetric KSVZ
axion model that the 1PI axino-gaugino-gauge boson amplitude at
energy scale in the range $M_\Phi < p < v_{PQ}$ is suppressed by
$M^2_\Phi/p^2$ in addition to the well-known suppression by
$p/16\pi^2 v_{PQ}$, where $M_\Phi$ is the supersymmetric mass of the
heaviest PQ-charged and gauge-charged matter field, which can be
well below $v_{PQ}$.
 On the other hand, the 1PI amplitude
 at energy scales below $M_\Phi$ has the same scaling behavior as the Wilsonian amplitude
(\ref{wil}), which can be noticed from (\ref{1pi2}), although the
precise coefficient can be different in general.
This observation applies also to another type of popular axion
model, the DFSZ model \cite{dfsz}. To see this, let us consider a
supersymmetric extension of the DFSZ axion model described by the
following form of the K\"ahler potential, gauge kinetic function and
superpotential at the UV scale $M_*$: \bea \label{dfsz1} K &=&
\sum_I \Phi_I^\dagger \Phi_I, \qquad f_a \, = \,
\frac{1}{\hat g_a^2(M_*)},\nonumber \\
W&=& h Z\left(XY-\frac{v_{PQ}^2}{2}\right)+\kappa
\frac{X^2}{M_*}H_uH_d+\cdots,\eea
where $\{\Phi_I\}$ denote the matter fields in the model, including
the MSSM Higgs, quark and lepton superfields, and the ellipsis
stands for the PQ-invariant Yukawa couplings for the quark and
lepton masses. The matter fields transform under the PQ symmetry as
\bea U(1)_{PQ}:\quad \Phi_I\rightarrow e^{ix_I\alpha} \Phi_I \eea
with the PQ charges  \bea
 \big(x_Z,x_X,x_Y,x_{H_u},x_{H_d}\big) = \big(0, -1, 1, 1
,1\big),\eea and the PQ charges of the  quark and lepton superfields
can be fixed by the MSSM Yukawa couplings. Here the Higgs $\mu$-term
is generated by the spontaneous breakdown of the PQ symmetry
\cite{kim_nilles,kim_nilles1}, yielding \bea \mu \, \sim \,
\frac{\kappa v_{PQ}^2}{2M_*}.\eea The most notable feature of this
DFSZ model is that  the Higgs doublets
 correspond to the heaviest PQ-charged  and gauge-charged matter field, so
$M_\Phi=\mu$ which should be around the weak scale. It is
straightforward to repeat our discussion for this DFSZ model, and
then one finds the 1PI axino-gaugino-gauge boson amplitude at
$p>\mu$ is suppressed by $\mu^2/p^2$. This would lead to a dramatic
reduction of the cosmological axino production, when compared to the
result of the previous analysis using  only the effective
interaction of the form (\ref{e11}). It should be noted that a
similar reduction is possible for the KSVZ model also as
$M_\Phi=M_Q$ can be in principle comparable to the weak scale.

Our result can have important implications for the cosmology of
supersymmetric axion model. For instance, certain parameter range of
the model and/or the cosmological scenario, which has been
considered to give a too large relic axino mass density in the
previous analysis,  can be safe if the model is assumed to have
$M_\Phi \ll v_{PQ}$. It also implies that at temperature $T>M_\Phi$,
axinos are produced dominantly by the processes involving the
heaviest PQ-charged  and gauge-charged matter supermultiplet, e.g.
${Q}+\mbox{gluon} \rightarrow \widetilde Q +\tilde a$ for the KSVZ model
and $\mbox{higgs}+W\rightarrow \mbox{higgsino} +\tilde a$ for the
DFSZ model, since the amplitude of such process does not involve the
loop suppression factor and is suppressed only by a single power of
$M_\Phi/v_{PQ}$. In the next section, we compute the axino
production rate and the resulting relic axino number density while
including the production by the heaviest PQ-charged and
gauge-charged matter multiplet  for both the KSVZ and DFSZ models.

It is in fact straightforward to generalize our discussion to
generic supersymmetric axion models. For this, let us consider a
general form of  Wilsonian effective lagrangian of the axion
superfield at a scale $\Lambda$ just below $v_{PQ}$, which takes the
form \bea \label{eff_general} {\cal L}_{\rm eff}(\Lambda)  &=& \int
d^4\theta \, \left( K_A(A+A^\dagger)+
Z_n(A+A^\dagger)\Phi_n^\dagger\Phi_n\right)\nonumber \\
 &+&\left[\, \int d^2\theta \, \left( \frac{1}{4} f^{\rm
eff}_a(A)W^{a\alpha}W^a_\alpha+W_{\rm eff}\right)+{\rm
h.c.}\,\right],\eea where $\{\Phi_n\}$ denote the light gauge-charged
matter fields  at $\Lambda$, and
 \bea \label{eff_general1} K_A &=& \frac{1}{2}(A+A^\dagger)^2
+{\cal O}\left(\frac{(A+A^\dagger)^3}{v_{PQ}}\right),
\nonumber \\
\ln Z_n&=& \left.\ln Z_n\right|_{A=0}+\tilde
y_n\frac{(A+A^\dagger)}{v_{PQ}}+{\cal
O}\left(\frac{(A+A^\dagger)^2}{v^2_{PQ}}\right),\nonumber \\
 f^{\rm eff}_a &=&
 \frac{1}{\hat g_a^2(\Lambda)}-\frac{C^a_W}{8\pi^2}\frac{A}{v_{PQ}},
\nonumber
\\
W_{\rm eff}&=& \frac{1}{2} e^{-(\tilde x_n+ \tilde
x_m)A/v_{PQ}}M_{mn}\Phi_m\Phi_n + \frac{1}{6}e^{-(\tilde x_n+\tilde
x_m+\tilde x_p)A/v_{PQ}}\lambda_{mnp}\Phi_m\Phi_n\Phi_p.\eea The PQ
symmetry in this effective theory is realized as \bea \label{pq3}
U(1)_{PQ}: \quad A\rightarrow A+i\alpha v_{PQ},\quad
\Phi_n\rightarrow e^{i \tilde x_n\alpha}\Phi_n,\eea and the
Wilsonian effective interactions of the axion superfield are given
by
 \bea
 \label{i1}
\Delta_1 {\cal L}(\Lambda) &=& -\int d^2\theta \,
\frac{C^a_W}{32\pi^2}\frac{A}{v_{PQ}}W^{a\alpha}W^a_\alpha,
 \\
 \label{i2}
\Delta_2 {\cal L}(\Lambda)&=&\int d^4\theta \,\, \tilde
y_n\frac{(A+A^\dagger)}{v_{PQ}} \Phi_n^\dagger \Phi_n,
 \\
\label{i3}\Delta_3 {\cal L}(\Lambda)&=&-\int d^2\theta
\,\frac{A}{v_{PQ}} \Big[\, \frac{1}{2}(\tilde x_m+\tilde x_n)M_{mn}
\Phi_n\Phi_m \nonumber \\
&& +\,\frac{1}{6}(\tilde x_m+\tilde x_n+\tilde
x_p)\lambda_{mnp}\Phi_m\Phi_n\Phi_p\,\Big].\eea According to our
discussion,  there are three quantities which are all related to the
axino coupling to gauge supermultiplets, but basically different
from each other:\bea \{\, C_W^a,\, C^a_{PQ},\, C^a_{1PI}\,\},\eea
where $C_W^a$ are the Wilsonian couplings  in (\ref{i1}), $C^a_{PQ}$
are the PQ anomaly coefficients defined as
 \bea
\partial_\mu
J^\mu_{PQ}=\frac{g^2}{16\pi^2} C_{PQ}^a F^{a\mu\nu}\widetilde
F^a_{\mu\nu},\eea
 and finally $C_{1PI}^a$ determines the leading part of the 1PI axino-gaugino-gauge boson amplitude\bea {\cal
A}^a_{1PI}(k,q,p) = -\frac{g^2}{16\pi^2 \sqrt{2}v_{PQ}}\widetilde
C^a_{1PI}\delta^4(k+q+p) \bar{u}(k)\sigma_{\mu\nu} \gamma_5
v(q)\epsilon^\mu p^\nu \eea which shows the behavior \bea
\label{1pi_def} && \widetilde C_{1PI}^a(k^2=q^2=0, \,M^2_{\rm light}< p^2 < M^2_{\rm heavy})\nonumber \\
& = & C_{1PI}^a+{\cal O}\left(\frac{M_{\rm
light}^2}{p^2}\ln^2\left(\frac{p^2}{M_{\rm light}^2}\right)\right)
+{\cal O}\left(\frac{p^2}{M_{\rm heavy}^2}\right) , \eea where
$M_{\rm light}$ and $M_{\rm heavy}$
 denote the masses of matter fields in the effective theory
(\ref{eff_general}).
 It is then straightforward to find \bea \label {pqanomaly}&& C^a_{PQ} \, =\,
C^a_W + 2\sum_n \tilde x_n {\rm
Tr}(T_a^2(\Phi_n)), \\
\label{c_1pi} && C_{1PI}^a(M_\Phi^2<p^2<\Lambda^2) \, =\,
C^a_W-2\sum_n \tilde y_n{\rm Tr}(T_a^2(\Phi_n)), \eea where
$M_{\Phi}$ is the mass of the heaviest PQ-charged and gauge-charged
matter field in the model. Here the expression of $C^a_{PQ}$ is
exact  and valid at any scale below $v_{PQ}$, while the expression
of $C^a_{1PI}$ is derived in 1-loop approximation. It is also
straightforward (see the Appendix A) to compute $C_{1PI}^a$ at lower
momentum scale,
which yields \bea C_{1PI}^a(\,p^2 < M_\Phi^2\,) &=&
C_W^a(\Lambda)-2\sum_{M^2_n < p^2} \tilde y_n(\Lambda) {\rm
Tr}(T_a^2(\Phi_n))\nonumber \\
 &+& 2\sum_{M_m^2>p^2}\tilde x_m(\Lambda)
{\rm Tr}(T_a^2(\Phi_m))\eea in 1-loop approximation.

In fact, one can easily derive an exact (in perturbation theory)
expression of $C_{1PI}^a$, including the piece depending
logarithmically on the momentum scale $p$. We already noticed that
$C_{1PI}^a$ can be determined by the real 1PI gauge kinetic function
 ${\cal F}_a$ as
\bea C_{1PI}^a = -8\pi^2 v_{PQ}\left.\frac{\partial {\cal
F}_a}{\partial A} \right|_{A=0}.\eea Solving the
Novikov-Shifman-Vainshtein-Zakharov RG equation  in the limit
$M_\Phi^2/p^2\rightarrow 0$,  one finds \cite{nsvz,arkani}
 \bea
\label{1pig} {\cal F}_a &=& {\rm Re}(f^{\rm eff}_a)
+\frac{b_a}{16\pi^2}\ln
\left(\frac{p^2}{\Lambda^2}\right)-\sum_n\frac{{\rm
Tr}(T_a^2(\Phi_n))}{8\pi^2}\ln {\cal Z}_n +\frac{{\rm
Tr}(T_a^2(G))}{8\pi^2}\ln {\cal F}_a,\eea where $b_a= -3{\rm
Tr}(T_a^2(G))+\sum_n{\rm Tr}(T_a^2(\Phi_n))$ is the one-loop beta
function coefficient, and
 $\gamma_n=d\ln {\cal
Z}_n/d\ln p^2$ is the anomalous dimension of $\Phi_n$ for the 1PI
wavefunction coefficient ${\cal Z}_n$ which can be chosen to satisfy
the matching condition \bea {\cal
Z}_n(p^2=\Lambda^2)=Z_n(\Lambda).\eea We then find \bea
C_{1PI}^a(M_\Phi^2 <p^2<\Lambda^2)=\frac{C^a_W-2\sum_n \tilde
y_n(p){\rm Tr}(T_a^2(\Phi_n))}{1-{\rm
Tr}(T_a^2(G))g_a^2(p)/8\pi^2},\eea  where \bea \tilde y_n(p ) =
v_{PQ}\left.\frac{\partial \ln {\cal Z}_n}{\partial
A}\right|_{A=0},\eea
 which gives (\ref{c_1pi})
at 1-loop approximation.



Within the effective theory (\ref{eff_general}), one can make a
holomorphic redefinition of matter fields \bea \label{redef} \Phi_n
\rightarrow e^{z_n A/v_{PQ}} \Phi_n,\eea after which the PQ symmetry takes
the form \bea U(1)_{PQ}:\quad A\rightarrow A+i\alpha v_{PQ}, \quad
\Phi_n\rightarrow e^{i(\tilde x_n-z_n)\alpha}\Phi_n,\eea and the
Wilsonian couplings of the axion superfield are changed as
 \bea
 \label{change}
 &&  C^a_W
\rightarrow C^a_W +2\sum_n z_n{\rm
Tr}(T_a^2(\Phi_n)),\nonumber \\
&& \tilde y_n \rightarrow \tilde y_n +z_n,\quad \tilde x_n
\rightarrow \tilde x_n-z_n.\eea This shows that one can always
choose a field basis with $\tilde x_n=0$, for which only the axion
superfield transforms under the PQ symmetry, and therefore
$C_W^a=C_{PQ}^a$. Another interesting choice would be the field
basis with $\tilde y_n=0$, for which $C_W^a=C_{1PI}^a$. Note that
$C_{PQ}^a$ and $C_{1PI}^a$ are directly linked to observables, and
therefore invariant under the reparametrization (\ref{change}) of
the Wilsonian couplings associated with the field redefinition
(\ref{redef}). It should be noticed also that for a given theory,
$C_{PQ}^a$ have common values at all scales, while $C_{1PI}^a$ can
have different values at different momentum scales.
 On the other hand, $C_W^a$ are lagrangian parameters which can take
different values in different field basis (or in different UV
regularization) within the same theory.

A key result of our discussion, which has direct implication for
cosmological axino production, is that the 1PI axino-gaugino-gauge
boson amplitude in the momentum range $M_\Phi^2 < p^2 < v_{PQ}^2$ is
suppressed by $M_\Phi^2/p^2$, more specifically\footnote{ This
is for the kinematic regime with a gauge boson (or gaugino)
4-momentum having $|p^2|>M_\Phi^2$, while the axino and gaugino (or
gauge boson) 4-momenta are on mass-shell. $\widetilde C_{1PI}^a$ can have
a bit different behavior in other kinematic regimes, but always
suppressed by $M_\Phi^2/p^2$ if any of the external particles has a
4-momentum $p$ with $|p^2|> M_\Phi^2$. For instance, if any of the
external particles has a vanishing 4-momentum, while the other
particles have $|p^2|>M_\Phi^2$, the amplitude is of ${\cal
O}\left(\frac{M_\Phi^2}{p^2}\ln (p^2/M_\Phi^2)\right)$.}
 \bea
\label{1pi_zero} \widetilde C_{1PI}^a(M_\Phi^2 < p^2
<v^2_{PQ})&=&C_{1PI}^a(M_\Phi^2 < p^2 <v^2_{PQ})+{\cal
O}\left(\frac{M_\Phi^2}{p^2}\ln^2\left(\frac{M_\Phi^2}{p^2}\right)\right)\nonumber
\\
&=&{\cal
O}\left(\frac{M_\Phi^2}{p^2}\ln^2\left(\frac{M_\Phi^2}{p^2}\right)\right),\eea
 where $M_\Phi$ is the supersymmetric mass of the heaviest
PQ-charged and gauge-charged matter field in the model, which can be
well below $v_{PQ}$ in most cases.
As we will see, there can be contributions to the above 1PI
amplitude from UV dynamics at scales above $v_{PQ}$, but they are
generically of ${\cal O}(M_\Phi^2\ln (M_*/v_{PQ})/8\pi^2v_{PQ}^2)$
or ${\cal O}(v_{PQ}^2/M_*^2)$, where $M_*$ is the fundamental
 scale
of the model, e.g. the Planck scale or the GUT scale, which is
presumed to be far above the PQ scale. These UV contributions are
either smaller than (\ref{1pi_zero}) or negligible by itself when
$M_*$ is comparable to the Planck or GUT scale.
As we will argue below, the result (\ref{1pi_zero}) applies to
generic supersymmetric axion model if the model has a UV realization
at $M_*$, in which (i) the PQ symmetry is linearly realized  in the
standard manner, \bea \label{uvpq} U(1)_{PQ}: \quad \Phi_I
\rightarrow e^{ix_I\alpha}\Phi_I,\eea where $\{\Phi_I\}$ stand for
generic chiral matter superfields, and (ii) all higher dimensional
operators of the model are suppressed by appropriate powers of
$1/M_*$.

To proceed, let $\{\Phi_A\}$ denote the gauge-singlet but
generically PQ-charged matter fields, whose VEVs break $U(1)_{PQ}$
spontaneously,  and $\{\Phi_n\}$ denote the gauge-charged matter
fields in the model.
 Then the K\"ahler
potential and superpotential at the UV scale $M_*$ can be expanded
in powers of the gauge-charged matter fields as follows \bea
K&=&K_{PQ}(\Phi_A^\dagger,\Phi_A)+\Big(1+ \frac{\kappa_{\bar AB
n}}{M_*^2}\Phi_A^\dagger \Phi_B+\cdots\Big)
\Phi_n^\dagger \Phi_n+\cdots, \nonumber \\
W&=& W_{PQ}(\Phi_A) +
\frac{1}{2}\Big(\hat\lambda_{Amn}\Phi_A+\frac{\hat
\lambda_{ABmn}}{M_*}\Phi_A\Phi_B+\cdots\Big)\Phi_m\Phi_n\nonumber \\
&&+ \, \frac{1}{6}\Big(\hat\lambda_{mnp}+
\frac{\hat\lambda_{Amnp}}{M_*}\Phi_A+\cdots\Big)\Phi_m\Phi_n\Phi_p+\cdots,\eea
where $K_{PQ}$ and $W_{PQ}$ are the K\"ahler potential and
superpotential of the PQ sector fields $\{\Phi_A\}$, and the
ellipses stand for higher dimensional terms. Here we assume that the
K\"ahler metric of the gauge-charged matter fields $\Phi_n$ is
flavor-diagonal for simplicity.  Also, to be complete, we includes
the leading higher dimensional operators suppressed by $1/M_*$.
Under the assumption that $K_{PQ}$ and $W_{PQ}$ provide a proper
dynamics to break the PQ symmetry spontaneously, we can parameterize
the PQ sector fields as follows \bea \Phi_A =
\left(\frac{1}{\sqrt{2}}v_A+ U_{Ai}\rho_i\right)e^{x_A
A/v_{PQ}},\eea where $v_A =\langle \Phi_A\rangle$ with
$v_{PQ}^2=\sum_A x_A^2 |v_A|^2$, $\rho_i$ denote the massive chiral
superfields in the PQ sector, and $U_{Ai}$ are the mixing
coefficients which are generically of order unity. For this
parametrization of the PQ sector fields, the K\"ahler potential and
superpotential at $M_*$ take the form \bea K &=&
K_{PQ}(\rho_i^\dagger, \rho_i, A+A^\dagger) + \left(Z_n^{(0)}+{\cal
O}\Big(\frac{v_{PQ}}{M_*^2}(A+A^\dagger),\frac{v_{PQ}\rho_i}{M_*^2},\frac{v_{PQ}\rho^\dagger_i}{M_*^2}\Big)\right)
 \Phi_n^\dagger \Phi_n+\cdots,
\nonumber \\
W &=& W_{PQ}(\rho_i) +\frac{1}{2} \left(M_{mn}+{\cal
O}\Big(\frac{M_{mn}\rho_i}{v_{PQ}}\Big)\right)e^{-(x_m+x_n)A/v_{pQ}}\Phi_m\Phi_n\nonumber
\\
&&+\, \frac{1}{6}\left(\lambda_{mnp}+{\cal
O}\Big(\frac{\rho_i}{M_*}\Big)\right)e^{-(x_m+x_n+x_p)A/v_{PQ}}\Phi_m\Phi_n\Phi_p+\cdots,\eea
where $Z_n^{(0)}$ and $M_{mn}$ are field-independent constants, and
 the Yukawa coupling constants $\lambda_{mnp}$ obeys the PQ selection rule \bea
 \label{pq_sel}
(x_m+x_n+x_p)\lambda_{mnp}=(x_m+x_n+x_p)\left(\hat\lambda_{mnp}+{\cal
O}\left(\frac{v_{PQ}}{M_*}\right)\right)={\cal
O}\left(\frac{v_{PQ}}{M_*}\right).\eea

One can now  integrate out the massive $\rho_i$ as well as the high
momentum modes of light fields, and also make the field redefinition
(\ref{redef}) to derive an effective theory in generic field basis.
The resulting effective lagrangian at the scale $\Lambda$ just below
$v_{PQ}$ takes the form of (\ref{eff_general1}) with \bea
\label{estimate} C^a_W &=& -8\pi^2 v_{PQ}\frac{\partial f_a^{\rm
eff}}{\partial A}= 2\sum_n z_n {\rm
Tr}(T_a^2(\Phi_n)), \nonumber \\
 \tilde x_n &=& x_n
-z_n,
\nonumber \\
\tilde y_n(\Lambda) &=& v_{PQ}\left.\frac{\partial \ln Z_n}{\partial
A}\right|_{A=0}= z_n + {\cal
O}\left(\frac{(x_m+x_n)}{8\pi^2}\frac{M_{mn}^2}{v_{PQ}^2}\ln
\left(\frac{M_*}{\Lambda}\right)\right)+{\cal O}\left(\frac{v_{PQ}^2}{M_*^2}\right)\nonumber \\
&=& z_n + {\cal O}\left(\frac{M_{\Phi}^2}{v_{PQ}^2}\right)+{\cal
O}\left(\frac{v_{PQ}^2}{M_*^2}\right),\eea where we have set  $\ln
(M_*/v_{PQ}) ={\cal O}(\pi^2)$ and used $(x_m+x_n)M_{mn}\lesssim
{\cal O}(M_\Phi)$ for $M_\Phi$ denoting the mass of the heaviest
PQ-charged and gauge-charged matter field in the model. Here $C_W^a$
and the first part of $\tilde y_n$ are due to the field redefinition
(\ref{redef}), the second part of $\tilde y_n$ is from the loops
involving the Yukawa couplings between the PQ sector fields and the
gauge-charged matter fields, which are generically of ${\cal
O}(M_{mn}/v_{PQ})$ and depend on the axion superfield through the
combination $(x_m+x_n)A$, and
 finally the last part
of $\tilde y_n$ is from the higher dimensional operator in the
K\"ahler potential of $\Phi_n$. There can be additional contribution
to $\tilde y_n$  from the loops involving the Yukawa couplings
$\lambda_{mnp}$ obeying the PQ selection rule (\ref{pq_sel}), which
is still within the estimate of $\tilde y_n$ in (\ref{estimate}). We
then have \bea \label{boun} C_{1PI}^a(p^2=\Lambda^2 ) &\, =\, &
C_{W}^a(\Lambda)-2\sum_n \tilde y_n(\Lambda){\rm Tr}(T_a^2(\Phi_n))
\nonumber \\
& =& {\cal O}\left(\frac{M_\Phi^2}{v_{PQ}^2}\right)+{\cal
O}\left(\frac{v_{PQ}^2}{M_*^2}\right)\eea at the cutoff scale
$\Lambda$ just below $v_{PQ}$, and the PQ selection rule
(\ref{pq_sel}) takes the form \bea (\tilde x_m +\tilde x_n +\tilde
x_p +\tilde y_m +\tilde y_n +\tilde y_p)\lambda_{mnp}= {\cal
O}\left(\frac{M_{\Phi}^2}{v_{PQ}^2}\right)+{\cal
O}\left(\frac{v_{PQ}}{M_*}\right).\eea

In the Appendix B, we examine the 1PI RG evolution of $C_{1PI}^a$
including higher loop effects, and show that the above estimate of
$C^a_{1PI}$ is valid at generic momentum scale in the range $M_\Phi
< p< v_{PQ}$. This implies that $\widetilde C_{1PI}^a$ in the momentum
range $M_\Phi < p < v_{PQ}$ is indeed dominated by the piece of
${\cal O}\left(\frac{M_\Phi^2}{p^2}\ln^2
\left(p^2/M^2_\Phi\right)\right)$, so the estimate (\ref{1pi_zero})
of $\widetilde C_{1PI}^a$ is valid even when higher loop effects are
taken into account.
 We thus conclude that in generic supersymmetric axion
model with a PQ scale hierarchically lower than the UV scale $M_*$,
which is presumed to be around the Planck scale or the GUT scale,
the 1PI axino-gaugino-gauge boson amplitude at momentum scales in
the range $M_\Phi < p < v_{PQ}$ is suppressed by $M_\Phi^2/p^2$, in
addition to the suppression by $p/16\pi^2 v_{PQ}$,  where $M_\Phi$
is the mass of the heaviest PQ-charged and gauge-charged matter
field in the model.
With the boundary condition (\ref{boun}) at $p^2=\Lambda^2$, one can
determine $C_{1PI}^a$ at lower momentum scale $p<M_\Phi$ by
computing the threshold correction. Using the result obtained in the
Appendix A, we find the leading constant part of $C_{1PI}^a$ at
generic momentum scale below $v_{PQ}$  is given by  \bea C_{1PI}^a(p)
&=& C_W^a(\Lambda)-2\sum_{M^2_n < p^2} \tilde y_n(\Lambda) {\rm
Tr}(T_a^2(\Phi_n)) +2\sum_{M_m^2>p^2}\tilde x_m(\Lambda) {\rm
Tr}(T_a^2(\Phi_m)) \nonumber \\
&=& 2\sum_{M^2_m>p^2} \Big(\tilde x_m(\Lambda)+\tilde
y_m(\Lambda)\Big) {\rm Tr}(T_a^2(\Phi_m)), \eea where $C_{W}^a(\Lambda),
\tilde y_n(\Lambda)$ and $\tilde x_n(\Lambda)$ are the Wilsonian
couplings in the effective lagrangian (\ref{eff_general}) at the
cutoff scale $\Lambda$ just below $v_{PQ}$. Note that this general
result correctly reproduces the 1PI amplitude (\ref{1pi2}) at
$p<M_Q$ in the KSVZ model.

\section{Thermal production of axino}

In this section, we examine the thermal production of axino with the
effective interactions which generically take the form
$(\ref{i1})-(\ref{i3})$. As we have noticed, if the model has a UV
completion (at a scale $M_*\gg v_{PQ}$) in which the PQ symmetry is
linearly realized in the standard manner and all non-renormalizable
interactions  are suppressed by the powers of $1/M_*$, the effective
interactions $(\ref{i1})-(\ref{i3})$ are constrained by the matching
condition (\ref{boun}) at the scale $\Lambda$ just below $v_{PQ}$.
Then one can choose a field basis in which the Wilsonian  couplings
of axion supermultiplet at $\Lambda$  are given by \bea
\label{basis1}C_W(\Lambda)&=& 0, \quad \tilde x_n \, = \,x_n, \quad
\tilde y_n(\Lambda)\,=\, {\cal O}\left(\frac{M_\Phi^2}{v_{PQ}^2},
\,\frac{v_{PQ}^2}{M_*^2}\right).
 \eea
 Of course, one can choose any other field
basis, for instance the one with \bea \label{basis2}C_W&=&2\sum_n
z_n {\rm Tr}(T_a^2(\Phi_n)),\quad \tilde x_n =x_n-z_n,\quad \tilde
y_n \,=\, z_n+ {\cal
O}\left(\frac{M_\Phi^2}{v_{PQ}^2},\,\frac{v_{PQ}^2}{M_*^2}\right),\eea
which would be obtained by the field redefinition
 (\ref{redef}).  Then the above Wilsonian couplings for arbitrary real
 values of $\{z_n\}$  should give the same physical results as
 those of (\ref{basis1}).
The field basis (\ref{basis1}) is convenient for describing the
physics at energy scales above $M_\Phi$ since the decoupling of the
axion supermultiplet in the limit $M_\Phi\rightarrow 0$ is manifest.
However, for physics at lower energy scales below $M_\Phi$, it is
often more convenient to choose the field basis (\ref{basis2}) with
$\tilde x_n =x_n-z_n=0$ for which $C_W=C_{PQ}$.

Let $\Phi,\Phi^c$ denote the heaviest PQ-charged and gauge-charged
matter superfield with a supersymmetric mass $M_\Phi$. In the field
basis (\ref{basis1}),  the relevant effective interaction of axion
supermultiplet takes a simple form \bea \label{eff} -\int d^2\theta
\, (x_\Phi+x_{\Phi^c}) M_\Phi\frac{A}{v_{PQ}} \Phi\Phi^c+{\rm
h.c.},\eea where we have ignored the small $\tilde y_n={\cal
O}(M_\Phi^2/v_{PQ}^2, v_{PQ}^2/M_*^2)$.
 In the KSVZ model (\ref{uv1}),
$\Phi,\Phi^c$ correspond to an exotic vector-like quark multiplet
with $x_\Phi+x_{\Phi^c}=1$, while in the DFSZ model (\ref{dfsz1}),
$\Phi,\Phi^c$ correspond to the Higgs doublet superfields $H_u,H_d$
with $M_\Phi=\mu$ and $x_\Phi+x_{\Phi^c}=2$.
 A key element for the
axino production by gauge supermultiplet is the 1PI
axino-gaugino-gauge boson amplitude which is given by
\begin{equation}
{\cal A}_{\text{1PI}}(k,q,p)=-\frac{g^2}{16\pi^2\sqrt{2}v_{PQ}}
\delta^4(k+q+p)\widetilde{C}_{\text{1PI}}(k,q,p)\bar{u}(k)\sigma_{\mu\nu}\gamma_5v(q)\epsilon^{\mu}(p)p^{\nu},
\label{axino:AWWcoupling}
\end{equation}
with
\begin{eqnarray}
\widetilde{C}_{\text{1PI}}(k^2=q^2=0; \, p^2\gg M_\Phi^2)
&\simeq&(x_\Phi+x_{\Phi^c})\frac{M_\Phi^2}{p^2}\ln^2\biggl(\frac{p^2}{M_\Phi^2}\biggr)
\label{1PI_coupling_1}\\
\widetilde{C}_{\text{1PI}}(k^2=q^2=0; \, p^2\ll M_\Phi^2)
&=&x_\Phi+x_{\Phi^c}+{\cal O}\biggl(\frac{p^2}{M_\Phi^2}\biggr).
\label{1PI_coupling_2}
\end{eqnarray}

\begin{figure}
\begin{center}
\subfigure[\label{proc_A_1}]{\includegraphics[width=3cm]{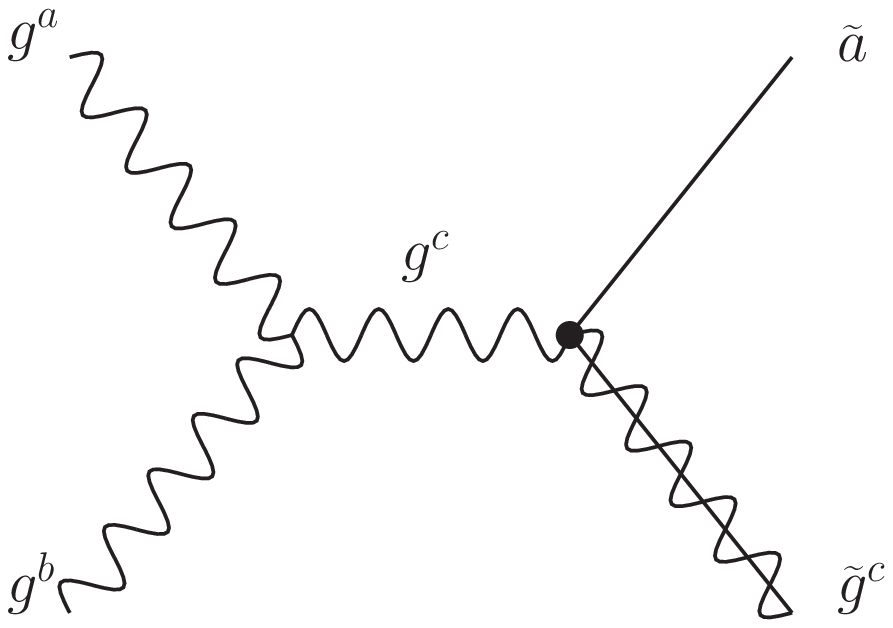}}
\quad
\subfigure[\label{proc_A_2}]{\includegraphics[width=3cm]{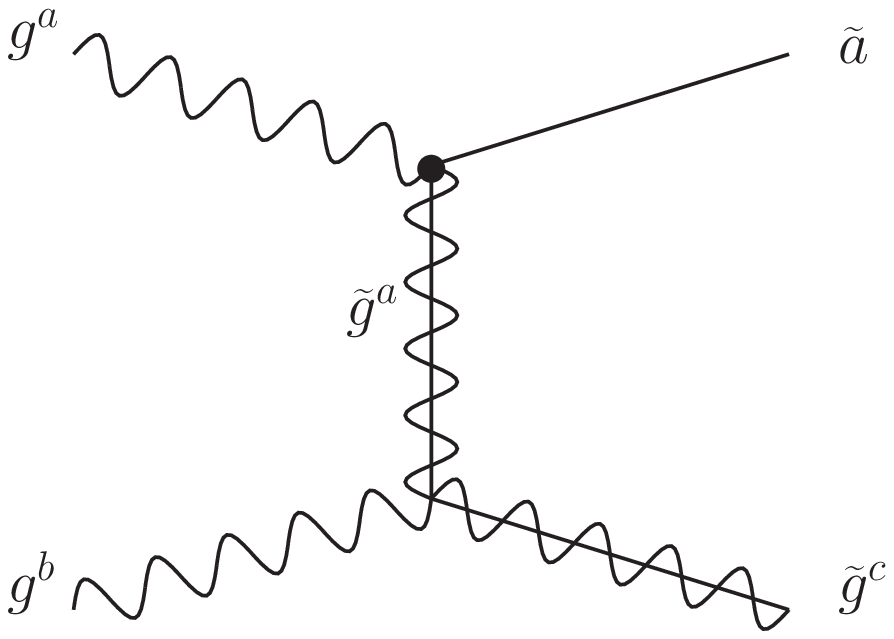}}
\quad
\subfigure[\label{proc_A_3}]{\includegraphics[width=3cm]{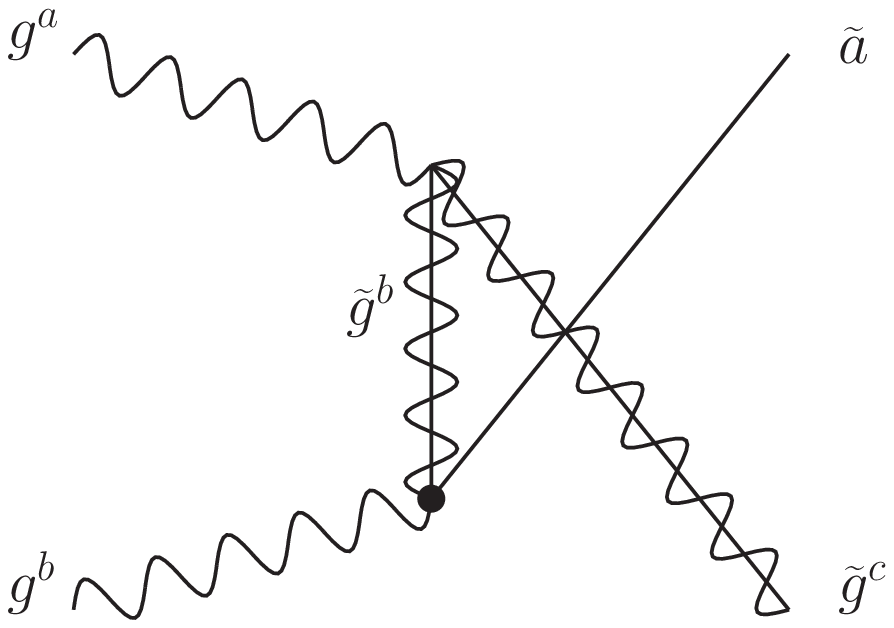}}
\quad
\subfigure[\label{proc_A_4}]{\includegraphics[width=3cm]{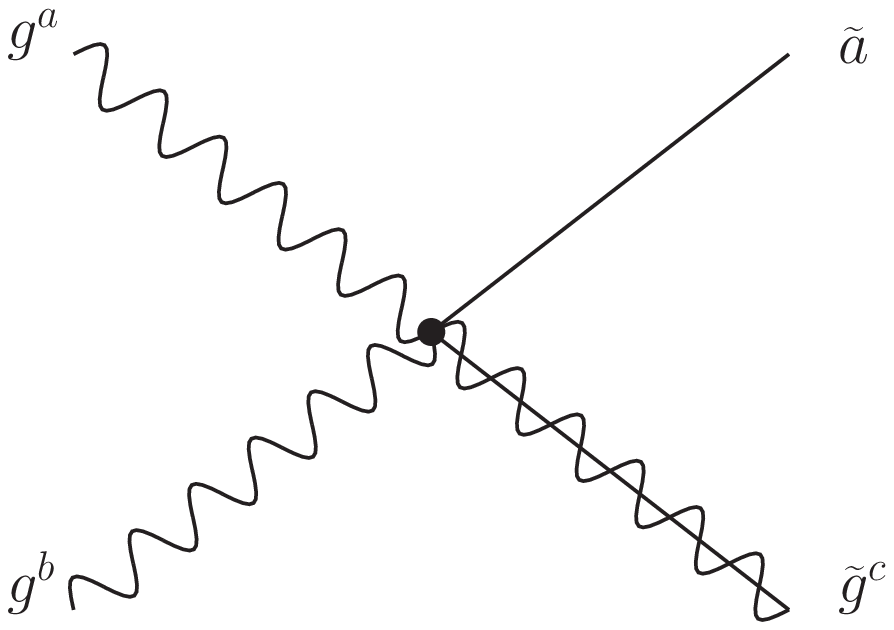}}
\caption{Diagrams for the process A. Here black dots represent the
1PI axino-gluino-gluon amplitude.\label{proc_A}}
\end{center}
\end{figure}

\begin{figure}
\begin{center}
\subfigure[\label{proc_C_1}]{\includegraphics[width=3cm]{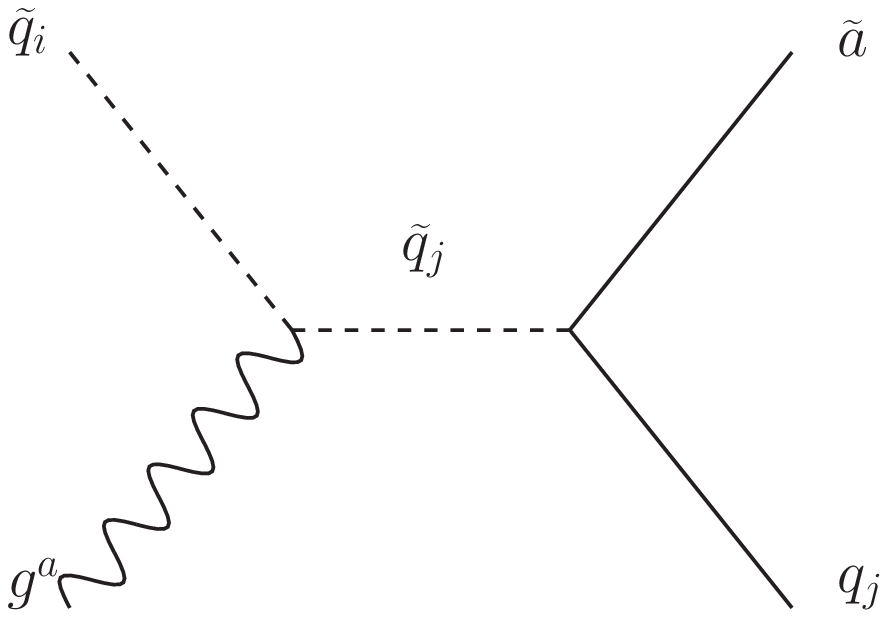}}
\quad
\subfigure[\label{proc_C_2}]{\includegraphics[width=3cm]{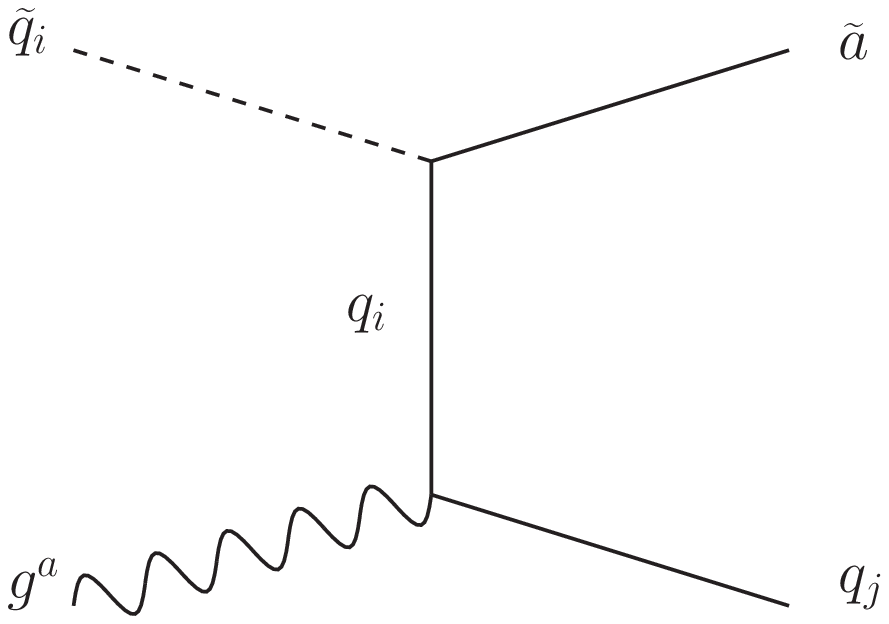}}
\quad
\subfigure[\label{proc_C_3}]{\includegraphics[width=3cm]{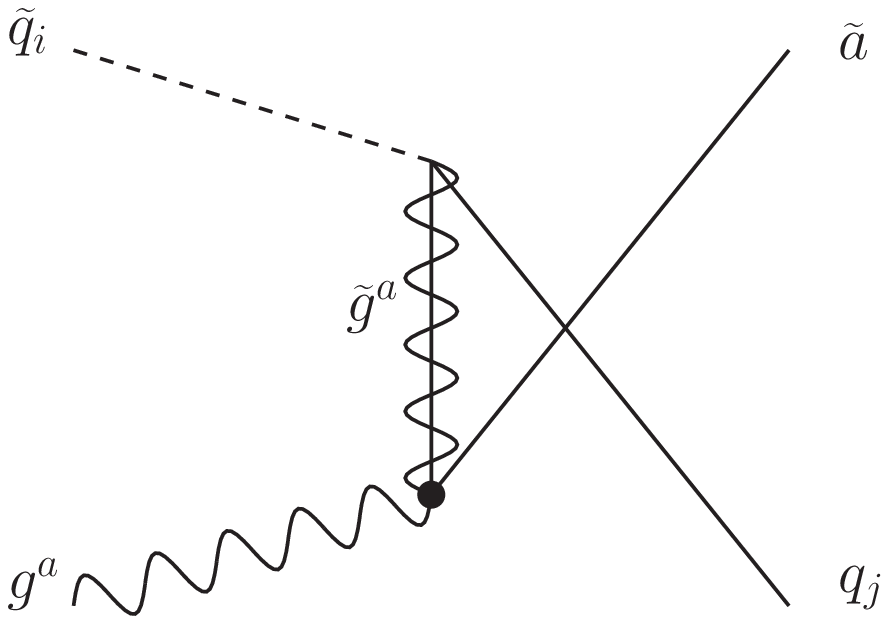}}
\caption{Diagrams for the process C. \label{proc_C}}
\end{center}
\end{figure}

\begin{figure}
\begin{center}
\subfigure[\label{proc_F_1}]{\includegraphics[width=3cm]{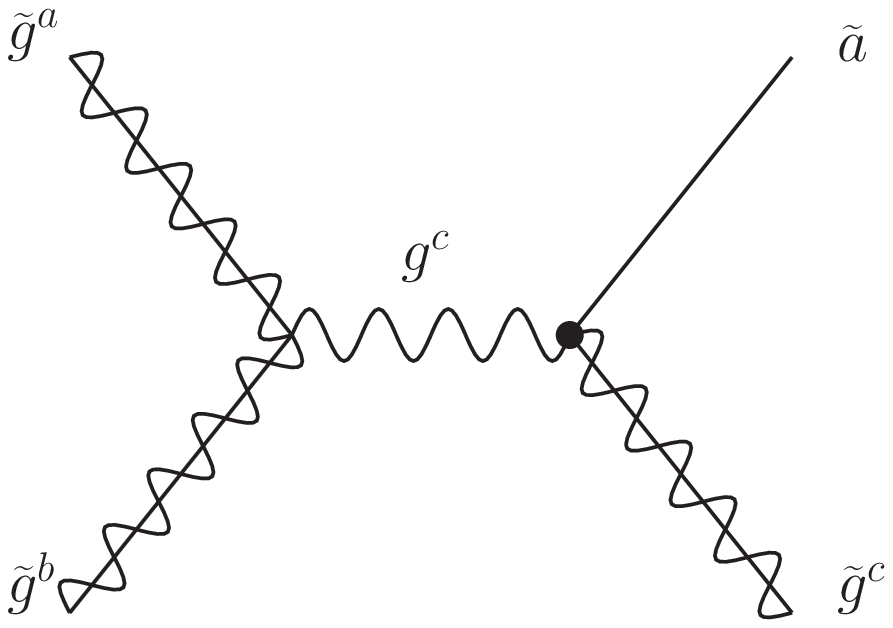}}
\quad
\subfigure[\label{proc_F_2}]{\includegraphics[width=3cm]{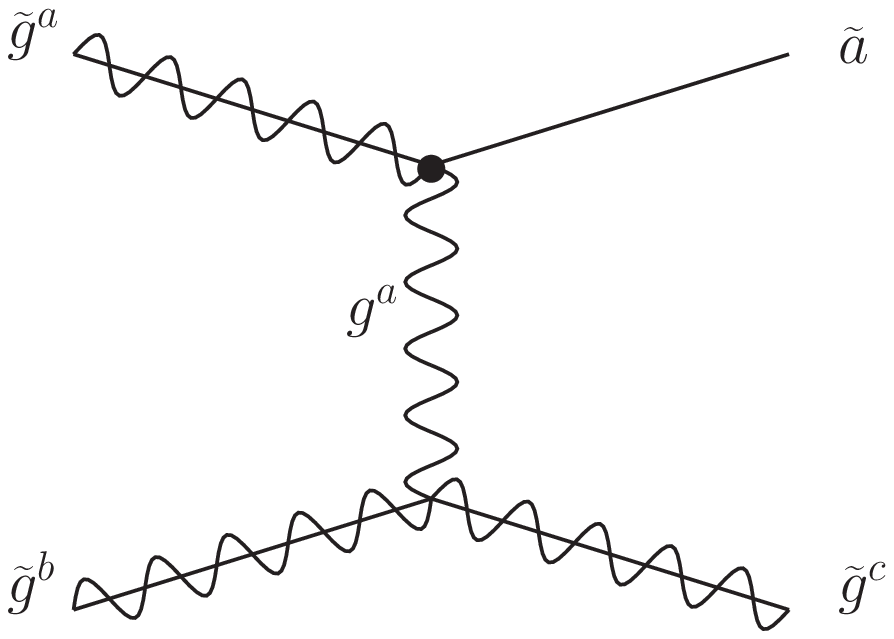}}
\quad
\subfigure[\label{proc_F_3}]{\includegraphics[width=3cm]{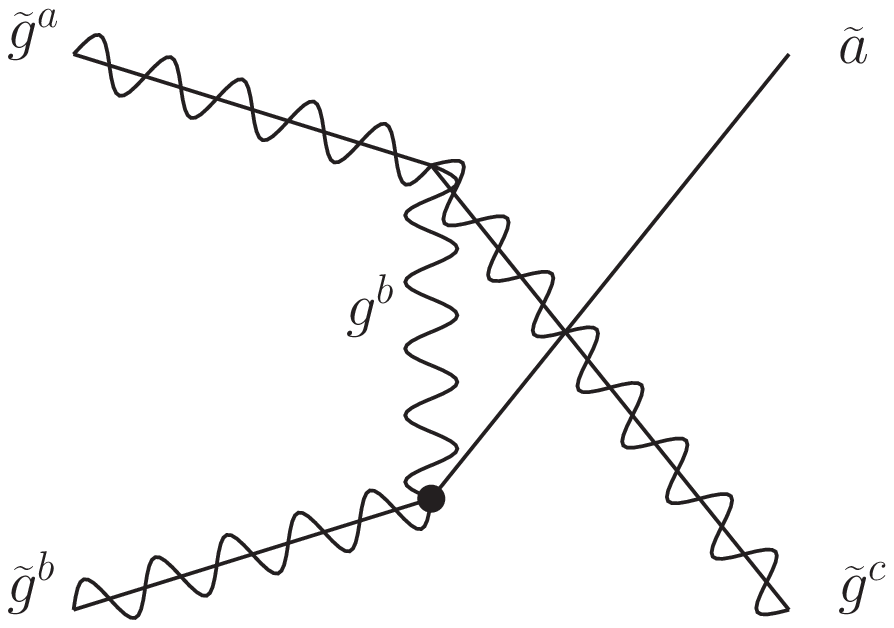}}
\caption{Diagrams for the process F.\label{proc_F}}
\end{center}
\end{figure}

\begin{figure}
\begin{center}
\subfigure[\label{proc_G_1}]{\includegraphics[width=3cm]{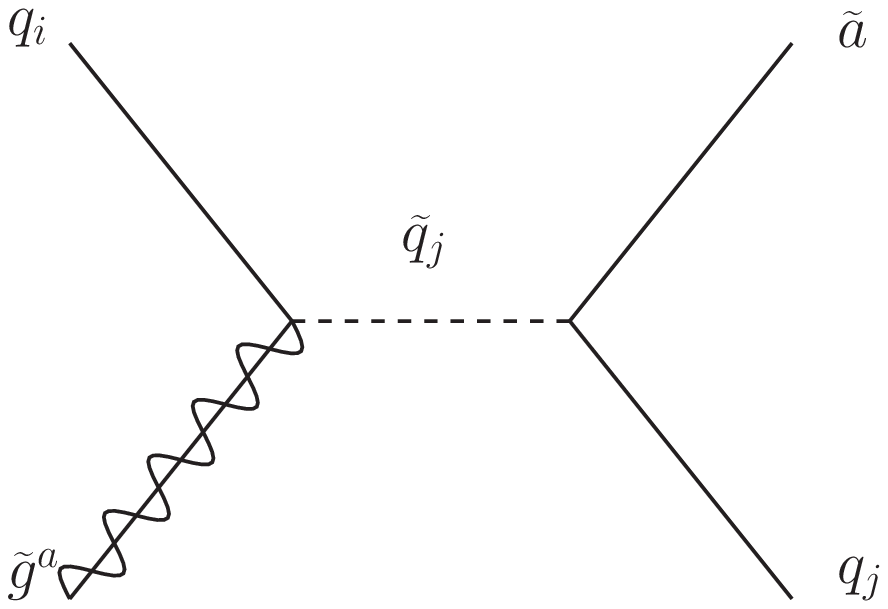}}
\quad
\subfigure[\label{proc_G_2}]{\includegraphics[width=3cm]{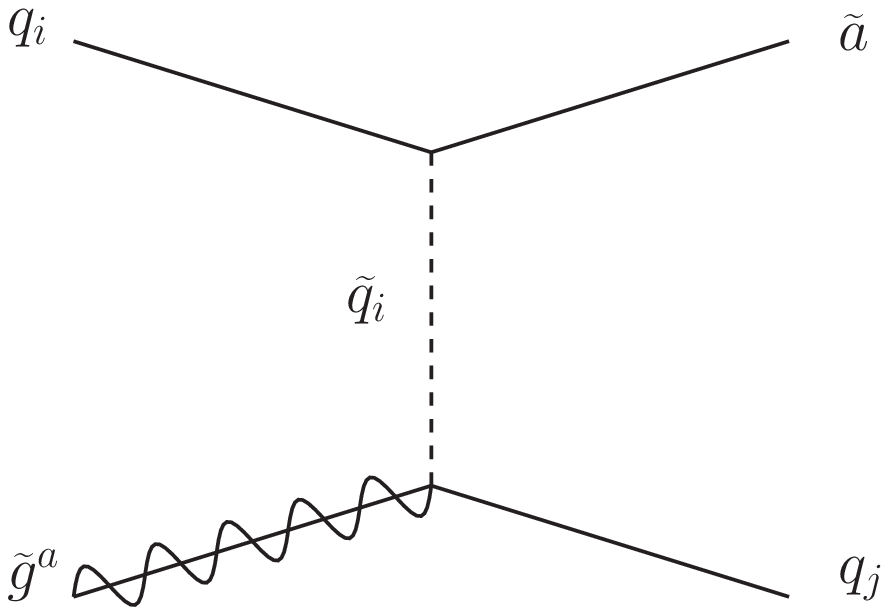}}
\quad
\subfigure[\label{proc_G_3}]{\includegraphics[width=3cm]{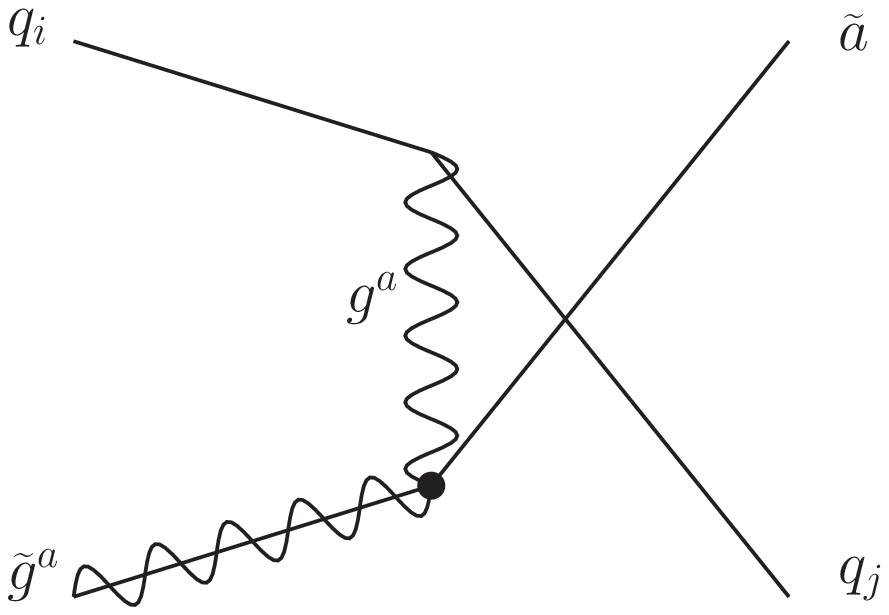}}
\caption{Diagrams for the process G.\label{proc_G}}
\end{center}
\end{figure}

\begin{figure}
\begin{center}
\subfigure[\label{proc_H_1}]{\includegraphics[width=3cm]{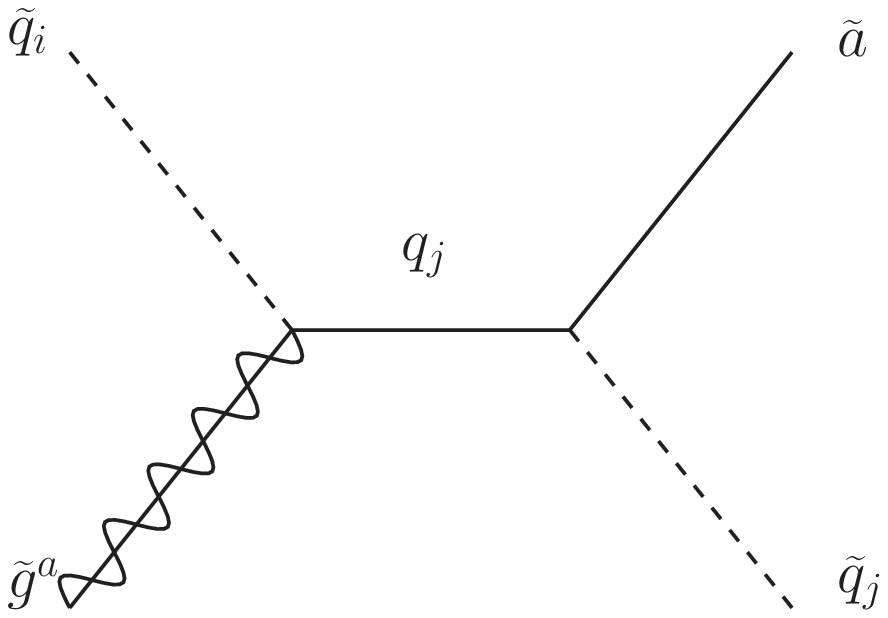}}
\quad
\subfigure[\label{proc_H_2}]{\includegraphics[width=3cm]{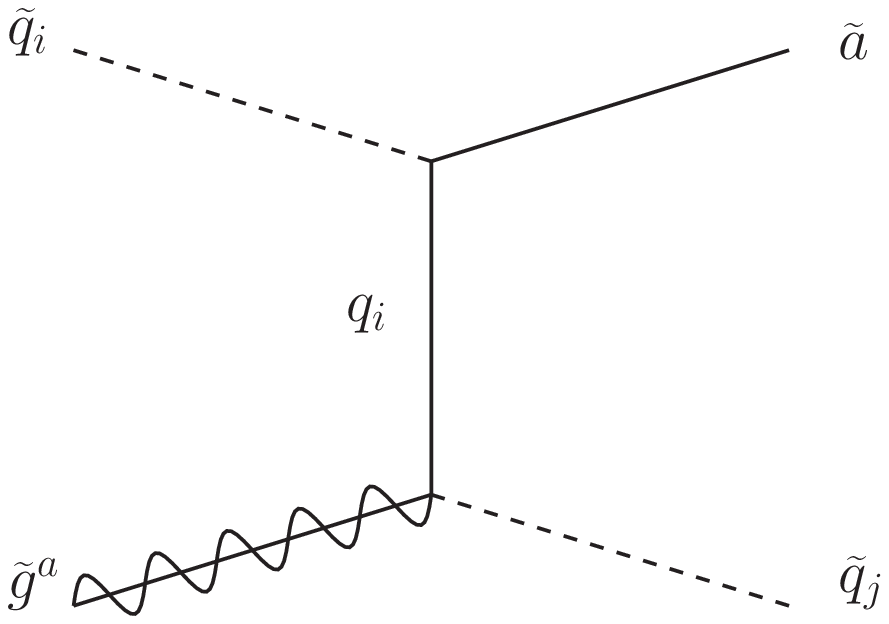}}
\quad
\subfigure[\label{proc_H_3}]{\includegraphics[width=3cm]{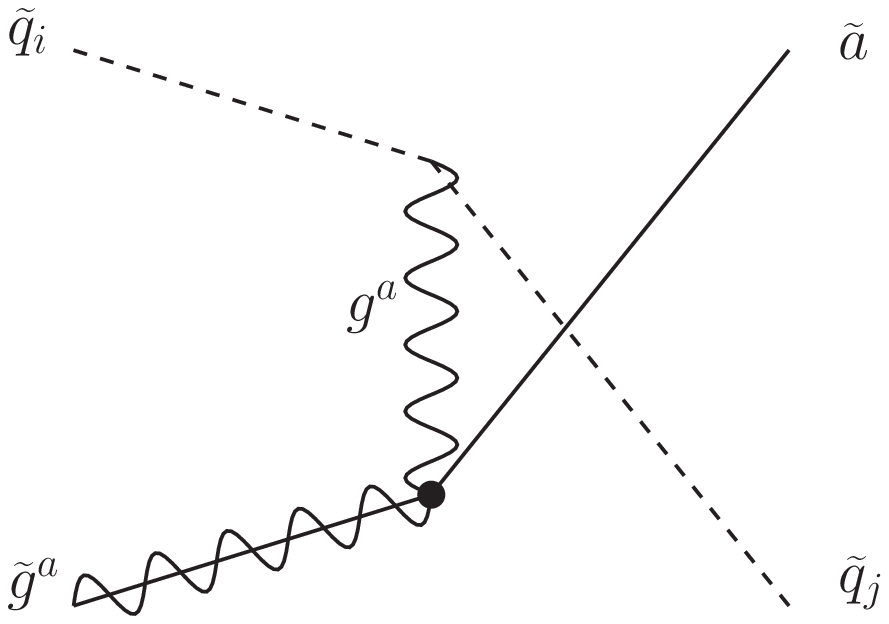}}
\quad
\subfigure[\label{proc_H_4}]{\includegraphics[width=3cm]{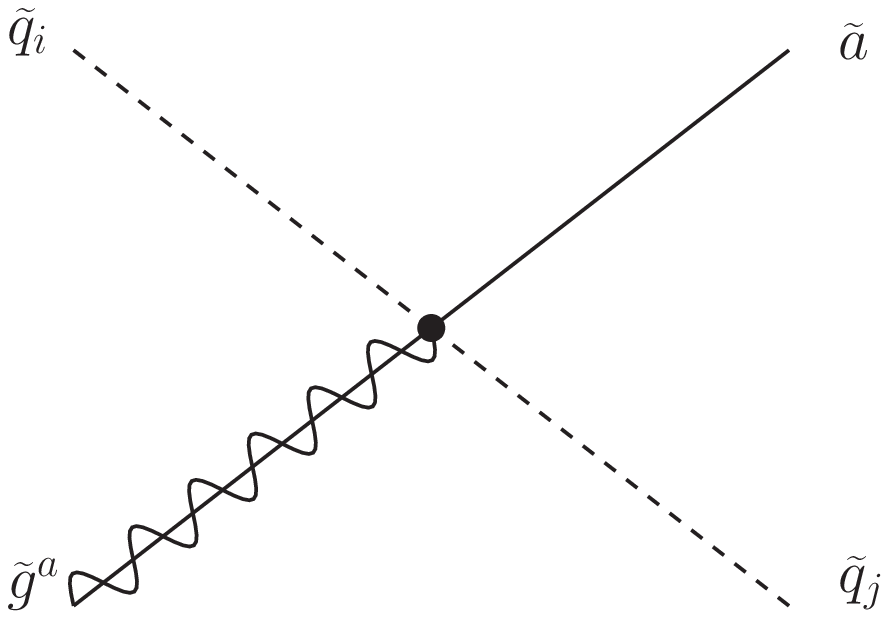}}
\caption{Diagrams for the process H.\label{proc_H}}
\end{center}
\end{figure}

With the above 1PI amplitude and also the axino-matter coupling
(\ref{eff}), we can calculate the thermal production of axinos in
the temperature range of our interest.   Following
\cite{Brandenburg}, here we consider the axino production processes
listed  in Table \ref{axino:proc}. (See  Fig. \ref{proc_A} $-$
\ref{proc_H} for corresponding Feynman diagrams.) Among these
processes, the processes A, B and F
produce axino through the 1-loop transition $g\rightarrow \tilde g
+\tilde a$ (or $\tilde g\rightarrow g+\tilde a$) whose amplitude is
given by the 1PI amplitude (\ref{axino:AWWcoupling}). On the other
hand, other processes
produce axino through both the tree-level transition
$\Phi\rightarrow \widetilde \Phi +\tilde a$ (or $\widetilde \Phi\rightarrow
\Phi+\tilde a$) and the 1-loop transition $g\rightarrow \tilde g
+\tilde a$ (or $\tilde g\rightarrow g+\tilde a$). To compute the
amplitudes of these axino production processes, we will use the
field basis  (\ref{basis1}) for which the decoupling of the axino in
the limit $M_\Phi\rightarrow 0$ is manifest.

The 1PI amplitude (\ref{1PI_coupling_1}) implies that the amplitude
of the axino production through the transition $g\rightarrow \tilde
g +\tilde a$ in the temperature range $M_\Phi\ll T < v_{PQ}$ is
suppressed by $M_\Phi^2/T^2$.  As a result, in this temperature
range, axinos are produced mostly by the transition $\Phi\rightarrow
\widetilde \Phi +\tilde a$ (or $\widetilde \Phi\rightarrow \Phi+\tilde a$)
with an amplitude ${\cal A}_{\Phi\widetilde \Phi\tilde
a}\propto(x_\Phi+x_{\Phi^c})M_\Phi/v_{PQ}$, and then the production
rate is given by \bea \Gamma_{\tilde a}(M_\Phi\ll T<v_{PQ})& =&
\sum_{IJ}\langle \sigma(I+J\rightarrow \tilde a+\cdots)v\rangle n_I n_J
\nonumber
\\
& = & {\cal O}(1)\times
(x_\Phi+x_{\Phi^c})^2\frac{g^2M_\Phi^2T^4}{\pi^5 v_{PQ}^2},
 \eea where $n_I$ is the number density of the $I$-th particle
 species in thermal equilibrium. On the other hand, at lower temperature
$T\ll M_\Phi$, the matter multiplet $\Phi$ is not available anymore,
and axinos are produced either by $g\rightarrow \tilde g +\tilde a$
or by $q \rightarrow \tilde q +\tilde a$, where $q$ denotes a
generic light matter multiplet with $M_q < T$. For the temperature
range $8\pi^2 M_q < T \ll M_\Phi$,
looking at the magnitudes  of the involved transition
amplitudes, one easily finds that axinos are produced mostly by the
transition $g\rightarrow \tilde g +\tilde a$ (or $\tilde
g\rightarrow g+\tilde a$) with an amplitude ${\cal A}_{g\tilde
g\tilde a}\propto (x_\Phi+x_{\Phi^c})/16\pi^2 v_{PQ}$, which results
in
\begin{equation}
\Gamma_{\tilde a}(8\pi^2M_q < T \ll M_\Phi) ={\cal O}(1) \times
(x_\Phi+x_{\Phi^c})^2\frac{g^6T^6}{64\pi^7v_{PQ}^2},
\end{equation}
where the ${\cal O}(1)$ factor includes the thermal field theoretic
effects discussed in \cite{Strumia}.

\begin{table}[tb]
\begin{center}
\begin{tabular}{c|c|c|c|c}
& ~Process~ & ~Feynman diagrams~ & ~$|{\cal M}|^2(8\pi^2 M_q \ll
T\ll M_\Phi)$~
& ~$|{\cal M}|^2(M_\Phi \ll T \ll v_{PQ})$~\\
\hline \hline ~A~ & ~$g+g\to\tilde{a}+\tilde{g}$~ & ~Fig.
\ref{proc_A}~
& ~$4C_1(s+2t+2t^2/s)$~ & ~suppressed~\\
~B~ & ~$g+\tilde{g}\to\tilde{a}+g$~ & ~crossing of A~
& ~$-4C_1(t+2s+2s^2/t)$~ & ~suppressed~\\
~C~ & ~$\tilde{q}+g\to\tilde{a}+q$~ & ~Fig. \ref{proc_C}~
& ~$2sC_2$~ & ~$-C\big(1+\frac{s-M_\Phi^2}{t-M_\Phi^2}\big)$~\\
~D~ & ~$q+g\to\tilde{a}+\tilde{q}$~ & ~crossing of C~
& ~$-2tC_2$~ & ~$C\big(1+\frac{t-M_\Phi^2}{s-M_\Phi^2}\big)$~\\
~E~ & ~$\overline{\tilde{q}}+q\to\tilde{a}+g$~ & ~crossing
of C~
& ~$-2tC_2$~ & ~$-C\frac{s-M_\Phi^2}{t-M_\Phi^2}$~\\
~F~ & ~$\tilde{g}+\tilde{g}\to\tilde{a}+\tilde{g}$~
& ~Fig. \ref{proc_F}~
& ~$-8C_1(s^2+t^2+u^2)^2/stu$~ & ~suppressed~\\
~G~ & ~$q+\tilde{g}\to\tilde{a}+q$~ & ~Fig. \ref{proc_G}~ &
~$-4C_2(s+s^2/t)$~
& ~$C\big(4+\frac{2M_\Phi^2}{s-M_\Phi^2}+\frac{2M_\Phi^2}{t-M_\Phi^2}\big)$~ \\
~H~ & ~$\tilde{q}+\tilde{g}\to\tilde{a}+\tilde{q}$~
& ~Fig. \ref{proc_H}~ & ~$-2C_2(t+2s+2s^2/t)$~
& ~$C\big(2-\frac{t-3M_\Phi^2}{s-M_\Phi^2}-\frac{s-3M_\Phi^2}{t-M_\Phi^2}\big)$~\\
~I~ & ~$q+\overline{q}\to\tilde{a}+\tilde{g}$~ & crossing of
G~ & ~$-4C_2(t+t^2/s)$~
& ~$C\big(4+\frac{2M_\Phi^2}{u-M_\Phi^2}+\frac{2M_\Phi^2}{t-M_\Phi^2}\big)$~\\
~J~ &
~$\tilde{q}+\overline{\tilde{q}}\to\tilde{a}+\tilde{g}$~
& ~crossing of H~ & ~$2C_2(s+2t+2t^2/s)$~ &
~$C\big(2-\frac{t-3M_\Phi^2}{u-M_\Phi^2}-\frac{u-3M_\Phi^2}{t-M_\Phi^2}\big)$~
\end{tabular}
\end{center}
\caption{Processes of axino production. Here $C=8g^2M_\Phi^2|T_{ij}(\Phi)^a|^2/v_{PQ}^2$, $C_1=g^6|f^{abc}|^2/256\pi^4v_{PQ}^2$,
and $C_2=g^6|T_{ij}^a(q)|^2/256\pi^4v_{PQ}^2$, where $\Phi$ denotes the
heaviest PQ-charged and gauge-charged matter multiplet with a
gauge-charge matrix given by $T^a_{ij}(\Phi)$.
 For $T\ll M_\Phi$, $q$ stands for generic gauge-charged
matter with $M_q < T$, while it means  $\Phi$ for $M_\Phi\ll T\ll
v_{PQ}$.
\label{axino:proc}}
\end{table}

Solving the Boltzmann equation, the relic axino number density over
the entropy density can be determined as (see the Appendix C)
\begin{equation}
Y_{\tilde a}(T_0)\equiv \frac{n_{\tilde
a}(T_0)}{s(T_0)}=\int_{T_0}^{T_R} \frac{dT}{T} \frac{\Gamma_{\tilde
a}}{s(T)H(T)}
\end{equation}
where $T_R$ is the reheat temperature, $s(T)=2\pi^2g_*T^3/45$ is the
entropy density, and $H(T)=\sqrt{\pi^2g_*/90}T^2/M_{Pl}$ is the
Hubble parameter for the effective degrees of freedom $g_*$ and the
reduced Planck mass $M_{Pl}=2.4\times10^{18}$ GeV. We then find \bea
Y_{\tilde a}( 8\pi^2 M_q < T_R \ll M_\Phi) &=& {\cal O}(1)\times
(x_\Phi+x_{\Phi^c})^2\frac{\bar gg^6M_{Pl}}{64\pi^7
v_{PQ}^2}T_R,\label{relic1}\\
Y_{\tilde a}(M_{\Phi}\ll T_R \ll v_{PQ}) &=& {\cal O}(1)\times
(x_\Phi+x_{\Phi^c})^2\frac{\bar gg^2M_{Pl}}{2\pi^4v_{PQ}^2}M_\Phi,
\label{relic2}\eea where $\bar g ={135\sqrt{10}}/{2\pi^3g_*^{3/2}}$.
Here we used the result of \cite{Strumia} for the first result,
while the second result is derived in the Appendix C.


\begin{figure}
\begin{center}
\includegraphics[width=8cm]{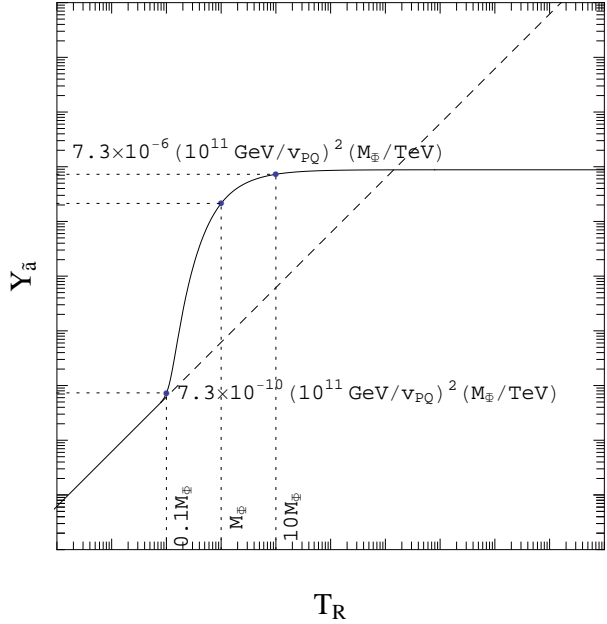}
\end{center}
\caption{Relic axino number density over the entropy density  vs the
reheating temperature $T_R$ (solid line). The dashed line represents
the result one would get by using only the effective interaction of
the form $\frac{1}{32\pi^2v_{PQ}}\int d^2\theta AW^aW^a$.
\label{result_general}}
\end{figure}

Fig. \ref{result_general} summarizes our discussion of the relic
axino density. It shows $Y_{\tilde a}\propto T_R$ for $T_R\lesssim
0.1 M_\Phi$, which is due to that in the temperature range $8\pi^2
M_q < T\lesssim 0.1 M_\Phi$, axinos are produced mostly through the
transition $g\rightarrow \tilde g+\tilde a$ (or $\tilde g\rightarrow
g+\tilde a$) with an amplitude ${\cal A}_{g\tilde g \tilde a}={\cal
O}(T/16\pi^2 v_{PQ})$.  (Here $M_q$ corresponds to the mass of the
next heaviest PQ-charged and gauge-charged matter multiplet.) If one
uses only the effective interaction of the form $\frac{1}{32\pi^2
v_{PQ}}\int d^2\theta AW^aW^a$  to evaluate the axino production by
$g\rightarrow \tilde g+\tilde a$ (or $\tilde g\rightarrow g+\tilde
a$), as one did in most of the previous analysis, one would get
$Y_{\tilde a}\propto T_R$ even for higher $T_R\gtrsim 0.1M_\Phi$, as
represented by the dashed line. Taking it into account that the 1PI
axino-gluino-gluon amplitude ${\cal A}_{g\tilde g\tilde a}={\cal
O}(M_\Phi^2\ln^2 (T/M_\Phi)/16\pi^2 Tv_{PQ})$\footnote{This estimate
is based on the simple identification  $p^2\sim T^2$, which might
not work for instance for the t-channel processes with the final
state axino momentum parallel to the initial state gluino (or gluon)
momentum \cite{Ellis:1984eq}. However, including the thermal mass of
intermediate state gluon (or gluino), which is of ${\cal O}(gT)$,
this simple approach does work for the purpose of the order of
magnitude estimation.} for $T> M_\Phi$, and therefore the axino
production at $T> M_\Phi$ is mostly due to the transition
$\Phi\rightarrow \widetilde \Phi+\tilde a$ (or $\widetilde \Phi\rightarrow
\Phi+\tilde a$) with an amplitude ${\cal A}_{\Phi\widetilde \Phi\tilde
a}={\cal O}(M_\Phi/v_{PQ})$, one can easily understand the behavior
of $Y_{\tilde a}$ for $T_R>10 M_\Phi$, which is nearly  independent
of $T_R$. Note that the dashed line crosses the correct solid line
at $T_R\sim 10^3 M_\Phi$, implying that the previous analysis based
on the effective interaction $\frac{1}{32\pi^2 v_{PQ}}\int d^2\theta
AW^aW^a$ alone gives rise to an overestimated axion relic density
for the reheat temperature $T_R \gtrsim 10^3 M_\Phi$, while it gives
an underestimated $Y_{\tilde a}$ for $0.1 M_\Phi\lesssim T_R\lesssim
10^3M_\Phi$.

We now proceed to apply our results to the two well-known types of
axion models, i.e. KSVZ \cite{ksvz} and DFSZ \cite{dfsz} models. In
DFSZ model, the heaviest PQ-charged and gauge-charged matter can be
identified as the MSSM Higgs doublets, and then $M_\Phi=\mu$ is
around the weak scale. On the other hand, in KSVZ model,
$\Phi,\Phi^c$ correspond to an exotic quark multiplet with $M_\Phi$
which can take any value between the weak scale and $v_{PQ}$. Since
our results can have more important cosmological implication when
$M_\Phi/v_{PQ}$ is smaller, we first discuss the case of DFSZ model.

\subsection{DFSZ axion model}

\begin{figure}
\begin{center}
\includegraphics[width=7cm]{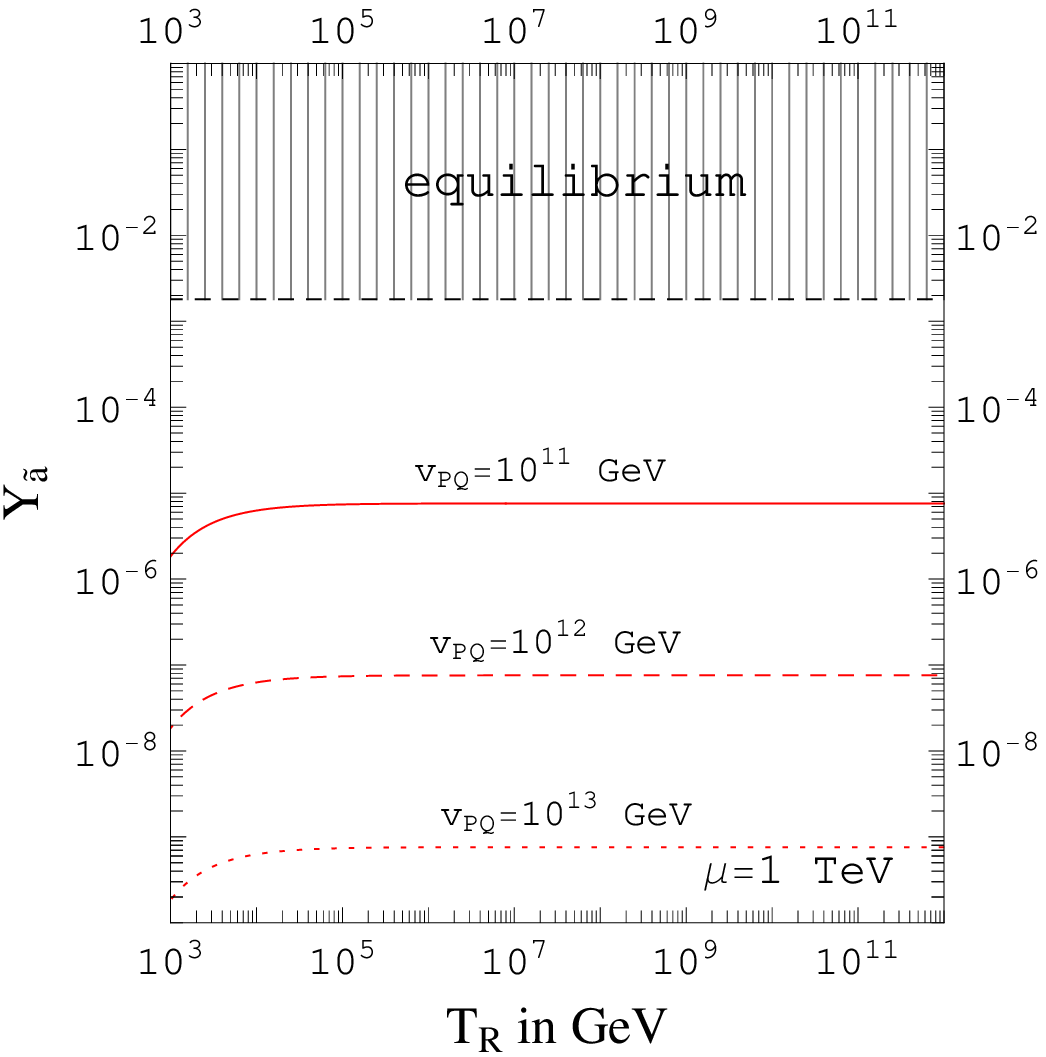}
\qquad
\includegraphics[width=7cm]{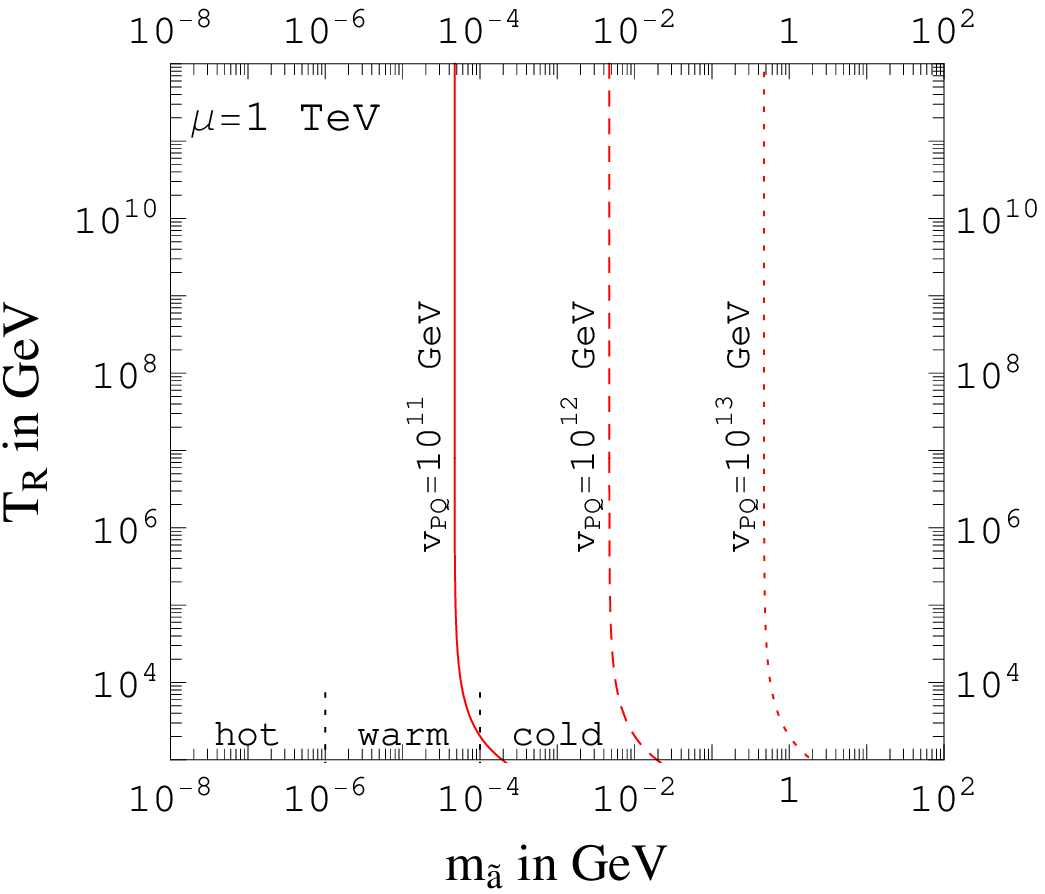}
\end{center}
\caption{(Left) Axino number density vs reheat temperature $T_R$ for
 $v_{PQ} = 10^{11}$ GeV (solid), $10^{12}$ GeV (dashed), and $10^{13}$ GeV (dotted).
 (Right) Contours giving $\Omega_{\tilde a}h^2 =0.1$. \label{result_DFSZ}}
\end{figure}

To apply our results to the DFSZ model in which $\Phi,\Phi^c$
correspond to the MSSM Higgs doublets $H_u,H_d$, we choose $M_\Phi =
\mu = 10^3$ GeV, and consider the axino production processes listed
in Table \ref{axino:proc} while identifying $(g,\tilde g)$ and
$(\Phi,\widetilde \Phi)$ as the $SU(2)_L$ gauge supermultiplet and the
Higgs supermultiplets, respectively.
 We use
$g^2=g_2^2(T=10^4 \,\,{\rm GeV}) \simeq 0.5$,  and the effective
degrees of freedom $g_*(T>\mu)=228.75$.
The results are  depicted in Fig. \ref{result_DFSZ}. In the left
panel, we plot $Y_{\tilde a}$ as a function of the reheat
temperature $T_R$ for three different PQ scales, $v_{PQ} = 10^{11}$
GeV (solid), $10^{12}$ GeV (dashed), and $10^{13}$ GeV (dotted). Our
result shows that  $Y_{\tilde a}$ has approximately a constant value
for $T_R >  10^4$ GeV, which is given by
\begin{equation}
Y_{\tilde{a}}(T_R> 10^4 \,\,{\rm GeV})\, \simeq \, 7.6 \times
10^{-8} \biggl(\frac{10^{12} \text{GeV}}{v_{PQ}}\biggr)^2.
\end{equation}
 In the figure,
the black dashed line represents  the axino number density when
axinos were in equilibrium with the thermal bath  of gauge-charged
particles, which is given by
\begin{equation}
Y^{\text{eq}}_{\tilde{a}}=\frac{0.42}{g_*} \simeq
1.8\times10^{-3}. \label{axino:equil}
\end{equation}
As we can see from Fig. \ref{result_DFSZ},  the axino interactions
with gauge-charged particles in DFSZ model are too weak for axinos
to be in thermal equilibrium even for high reheat temperature
comparable to $v_{PQ}$.

In the right panel of Fig. \ref{result_DFSZ}, we plot the contours
giving the relic axino abundance which would explain the observed DM
density $\Omega_{\rm DM} h^2\simeq 0.1$  \cite{cdm}.
As an approximate guideline, we can refer to hot, warm, and cold
axino dark matter for axino mass in the range $m_{\tilde{a}}
\lesssim 1$ keV, 1 keV $\lesssim m_{\tilde{a}} \lesssim 100$
keV, and $m_{\tilde{a}} \gtrsim 100$ keV, respectively
\cite{Brandenburg}. Then as shown in the figure, if $v_{PQ}\gtrsim
\mbox{few}\times 10^{11}$ GeV, axino can provide the {\em cold} dark
matter in our Universe with correct relic abundance. In our case,
there is no upper bound on the reheat temperature coming from the
relic dark matter density, and therefore
we can avoid the previous conclusion that cold axino dark matter
scenario is viable only when the reheat temperature is relatively
low as $T_R\lesssim 10^6$ GeV \cite{Brandenburg}. In this analysis,
we have ignored the effects of SUSY breaking and electroweak
symmetry breaking on the axino effective interactions, which should
be a good approximation for $T\gg 10^3$ GeV. We note that the
Higgsino (or the Higgs boson) decay into axino  can be sizable if
one includes SUSY breaking effect around the weak scale. If one
includes the electroweak symmetry breaking effect, stop decay also
can give a sizable contribution \cite{axino_decay2}. At any rate, as
long as $T_R>M_\Phi$,  the axino production by the decay processes
does not significantly alter our result depicted in Fig.
\ref{result_DFSZ}.

For axinos to be a successful cold dark matter, we should also
consider the constraints from the big bang nucleosynthesis (BBN).
The long-lived next lightest SUSY particle (NLSP) such as the
bino-like or wino-like lightest neutralino $\chi$ or the lighter
stau ${\tilde \tau}$ can be problematic in BBN if it destroys
the light primordial elements by emitting an energetic photon or
hadronic shower through the decay into axino. Neutralino decay to
gauge boson and axino, $\tilde{\chi}\to\gamma/Z+\tilde{a}$
can be induced by the 1PI axino-gaugino-gauge boson amplitude
(\ref{axino:AWWcoupling}), and its rate is estimated as
 \bea
 \label{decay1}
\Gamma(\tilde{\chi}\to\gamma/Z+\tilde{a}) \,\sim \,
\frac{1}{16\pi}\biggl(\frac{g_2^2}{8\pi^2}\biggr)^2\frac{M_{\tilde{\chi}}^3}{v_{PQ}^2}\,
\sim\,\frac{1}{0.1\text{ s}}
\biggl(\frac{M_{\tilde{\chi}}}{200\text{ GeV}}\biggr)^3
\biggl(\frac{10^{12}\text{ GeV}}{v_{PQ}}\biggr)^2, \eea where $g_2$
is the $SU(2)_L$ coupling constant. On the other hand, the stau
decay  $\tilde{\tau}\to\tau+\tilde{a}$ is induced by
tree-level process in DFSZ model since there is axino-gaugino mixing due to
the electroweak symmetry breaking, and  its decay rate can be estimated as \cite{stau_decay_dfsz}
\begin{equation}
\label{decay2} \Gamma(\tilde{\tau}\to\tau+\tilde{a})
\,\sim\,\frac{m_{\tilde{\tau}}}{16\pi}\frac{g_2^2v_{\textrm{weak}}^2}{v_{PQ}^2}
\frac{m_Z^2}{M_2^2}\cos^2{\theta_W}\sin^4{\beta}
\,\sim\,\frac{1}{10^{-5}\text{
s}}\biggl(\frac{m_{\tilde{\tau}}}{200\text{ GeV}}\biggr)
\biggl(\frac{10^{12}\text{ GeV}}{v_{PQ}}\biggr)^2.
\end{equation}
Hence, the NLSP neutralino lifetime is ${\cal O}(10^{-7})-{\cal
O}(10^{-1})$ s, while the NLSP stau lifetime is ${\cal
O}(10^{-11})-{\cal O}(10^{-5})$ s for $v_{PQ}=10^9-10^{12}$ GeV. In
order to avoid the BBN constraints, the lifetime of such long-lived
particles is required to be shorter than $10^2$ s
\cite{Feng:2004zu}, which can be easily satisfied in DFSZ model over
a reasonable parameter range of the model.

\subsection{KSVZ axion model}

\begin{figure}
\begin{center}
\includegraphics[width=7cm]{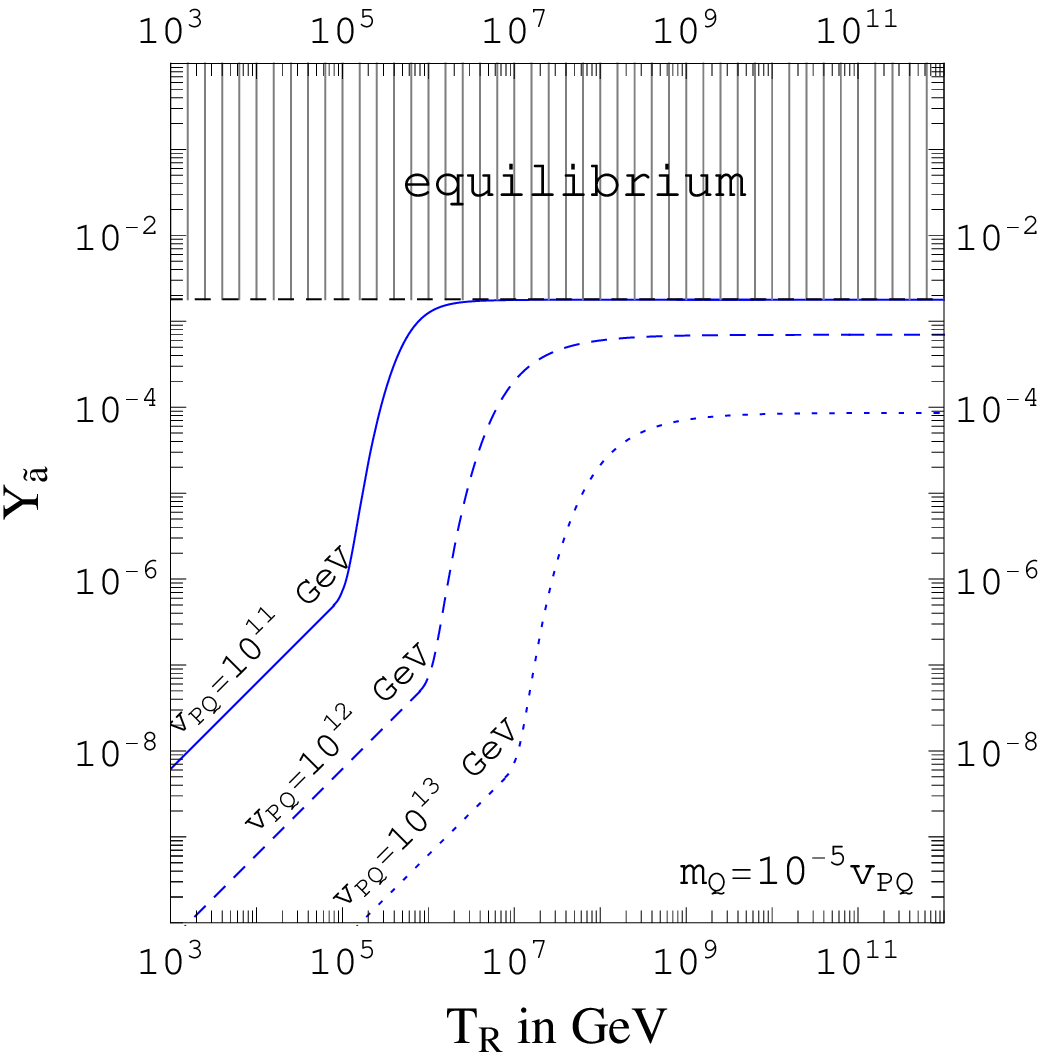}
\qquad
\includegraphics[width=7cm]{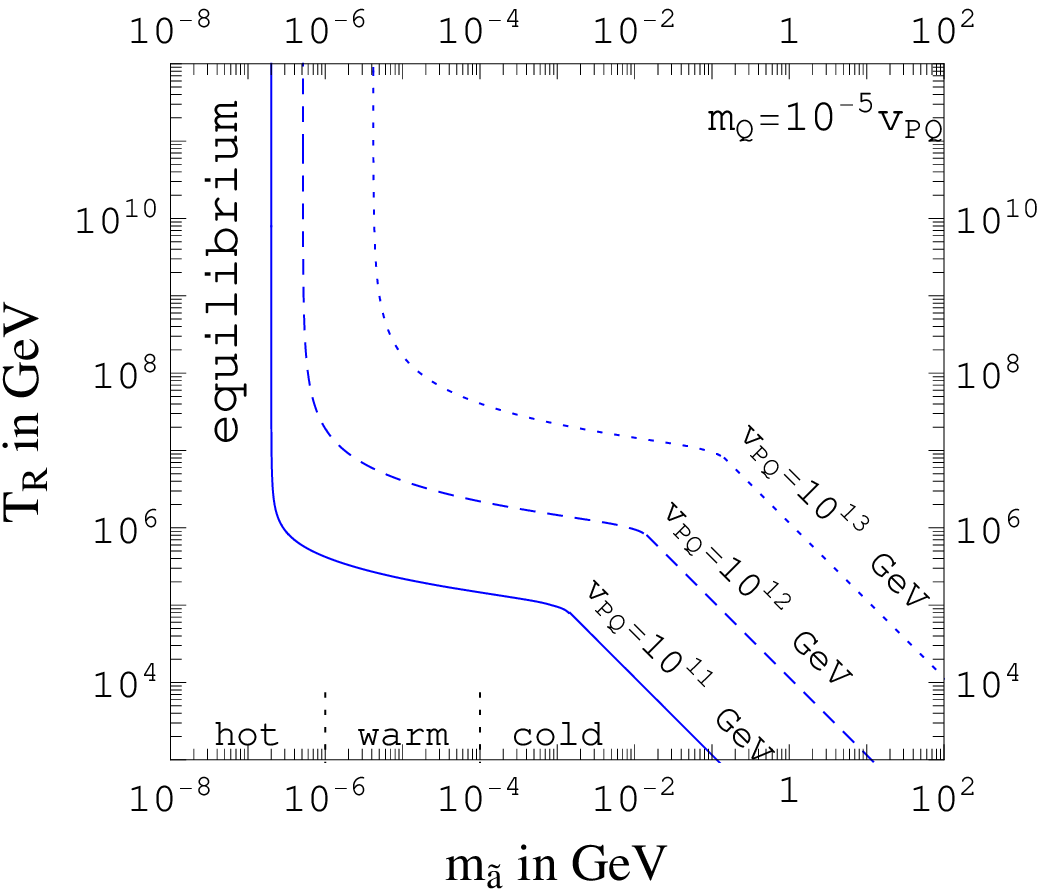}
\end{center}
\caption{(Left) Axino number density vs reheat temperature $T_R$ for
$v_{PQ} = 10^{11}$ GeV (solid), $10^{12}$ GeV (dashed), and
$10^{13}$ GeV (dotted) with $M_Q/v_{PQ}=10^{-5}$.
(Right) Contours giving $\Omega_{\tilde a}h^2 =0.1$.}
\label{result_KSVZ}
\end{figure}

Let us now consider the KSVZ axion model in which the heaviest
PQ-charged and gauge-charged field is an exotic quark multiplet
$Q,Q^c$ with $M_Q$ which can take any value between the weak scale
and the PQ scale. Fig. \ref{result_general} suggests that the
previous analysis of axino production based on the effective
interaction $\frac{1}{32\pi^2 v_{PQ}}\int d^2\theta AW^aW^a$ alone
can be applied only for $T_R\sim 0.1 M_Q$. More specifically, axinos
are produced mostly by the gluon supermultiplet at $T\lesssim 0.1
M_Q$, while at higher temperature axino production is mostly due to
the transition $Q\rightarrow \widetilde Q+\tilde a$ (or $\widetilde
Q\rightarrow Q+\tilde a$). Thus our discussion in this paper can
significantly alter the previous result if $M_Q$ is well below
$v_{PQ}$. To be specific, here we consider $M_Q = 10^{-5} v_{PQ}$,
and depict the resulting relic axino density in the left panel of
Fig. \ref{result_KSVZ}
 for $T_R >10^3$ GeV and  $v_{PQ} = 10^{11}$ GeV (solid), $10^{12}$ GeV
(dashed), and $10^{13}$ GeV (dotted). For numerical result, we use
$g^2=g_3^2(T=10^6 - 10^8\,\, {\rm GeV}) \simeq 1$, and find \bea
Y_{\tilde{a}}(T_R\gg M_Q) &\simeq& 8.8 \times 10^{-4}
 \biggl(\frac{10^{12}
\text{GeV}}{v_{PQ}}\biggr)^2\biggl(\frac{M_Q}{10^7 \text{GeV}}\biggr), \nonumber \\
Y_{\tilde a}(T\lesssim 0.1 M_Q) &\simeq & 2.2 \times 10^{-6}
\biggl(\frac{10^{12} \text{GeV}}{v_{PQ}}\biggr)^2
\left(\frac{T_R}{10^7{\rm GeV}}\right).
 \eea   Again, the black dashed line in the figure
represents  the axino number density when axinos were in equilibrium
with the thermal bath of gauge-charged particles.
 Our result shows that axinos could indeed be in thermal equilibrium if
 $v_{PQ} < 10^{12}$ GeV and the reheat temperature is high enough.

  In the right panel of Fig. \ref{result_KSVZ}, we plot the
contours giving $\Omega_{\tilde a}h^2\simeq 0.1$, which shows that
 the CDM constraint gives a severe upper bound on the reheat temperature
 depending upon the axino mass.
Note that our analysis includes the axino productions by the exotic
quark multiplet $Q,Q^c$, which in fact provide dominant production
channels for $T\gtrsim 0.1 M_Q$. As a result, for certain range of
the axino mass, the upper bound on $T_R$ can be more stringent than
the bound one would obtain based on the effective interaction
$\frac{1}{32\pi^2 v_{PQ}}\int d^2\theta AW^aW^a$ alone. Still, one
can relieve the bound by assuming that $M_Q$ is small enough, e.g.
close to the weak scale, for which the situation becomes similar to
the case of DFSZ model.

As in DFSZ model, we should consider the BBN constraint on the
decays of long-lived NLSP.
 For the neutralino decays $\tilde{\chi}\to\gamma/Z+\tilde{a}$, the rate is
 given by (\ref{decay1})
as in DFSZ model. However, there is no tree level coupling for the
stau decay $\tilde{\tau}\to\tau+\tilde{a}$
 in KSVZ model, and the decay rate is suppressed by the two-loop factor
 as \cite{axino_decay2}
\bea \Gamma(\tilde{\tau}\to\tau+\tilde{a}) &\sim&
\frac{1}{16\pi}\biggl(\frac{9\sqrt{2}\alpha_{\rm
em}^2e_Q^2}{8\pi^2\cos^4\theta_W}\biggr)^2
\ln^2\biggl(\frac{M_Q^2}{m_{\tilde{\tau}}^2}\biggr)
\biggl(\frac{m_{\tilde{\tau}}^3}{2v_{PQ}^2}\biggr)\nonumber \\
&\sim& \frac{1}{10^4\text{ s}}
\biggl(\frac{m_{\tilde{\tau}}}{200\text{ GeV}}\biggr)^3
\biggl(\frac{10^{12}\text{ GeV}}{v_{PQ}}\biggr)^2, \eea where
$\alpha_{\rm em}$ is the fine structure constant and $e_Q=1/3$ is
the $U(1)_{\text{em}}$ charge of $Q$. Hence, neutralino lifetime is
the same as in the DFSZ case, while stau lifetime is ${\cal
O}(10^{-2})-{\cal O}(10^4)$ s for $v_{PQ}=10^9-10^{12}$ GeV.
Therefore, the neutralino NLSP is safe as before, while the stau
NLSP is safe only for a relatively heavy $m_{\tilde{\tau}}$.

\section{Conclusion}

In this paper, we have discussed certain features of the effective
interactions of axion supermultiplet, which might be important for
the cosmology of supersymmetric axion model, and examined the
implication to the thermal production of axinos in the early
Universe. If the model has a UV completion at a fundamental scale
$M_*\gg v_{PQ}$, in which the PQ symmetry is linearly realized in
the standard manner and all non-renormalizable interactions are
suppressed by the powers of $1/M_*$, the axion supermultiplet is
decoupled from the gauge-charged fields  in the limit
$M_\Phi/v_{PQ}\rightarrow 0$ and $v_{PQ}/M_*\rightarrow 0$, where
$M_\Phi$ is the mass of the heaviest PQ-charged and gauge-charged
matter multiplet in the model.  As a result, in models with small
values of $M_\Phi/v_{PQ}$ and $v_{PQ}/M_*$, the axino production
rate at temperature $T\gg M_\Phi$ is suppressed by the powers of
small $M_\Phi/T$. Such decoupling feature is not manifest in generic
form of the effective lagrangian of axion supermultiplet, however it
should be imposed as a matching condition at the scale just below
$v_{PQ}$.

Our observation is particularly important for the cosmology of
supersymmetric DFSZ axion model in which $M_\Phi$ corresponds to the
MSSM Higgs $\mu$-parameter which is far below the PQ scale $v_{PQ}$.
Cosmology of KSVZ axion model can be significantly altered also, if
the PQ-charged exotic quark has a mass well below $v_{PQ}$. We have
performed an explicit analysis of the thermal production of axinos
for both the DFSZ and KSVZ axion models,  and presented the
resulting relic axino density as well as the bound on the reheat
temperature $T_R$ coming from the observed dark matter density in
our Universe. Our analysis does not take into account the effects of
SUSY breaking and electroweak symmetry breaking. For $T_R$ close to
the weak scale, a more careful analysis is required, including the
NLSP decays into axino as well as the mixing due to SUSY breaking
and electroweak symmetry breaking, and it will be the subject of
future work \cite{future}.

\section*{Acknowledgement}

We thank L. Covi and E. J. Chun for useful discussions.
KC and SHI are supported by the KRF Grants funded by the Korean
Government (KRF-2008-314-C00064 and KRF-2007-341-C00010) and the
KOSEF Grant funded by the Korean Government (No. 2009-0080844).
 KJB
is supported by TJ Park Postdoctoral Fellowship of POSCO TJ Park
Foundation. We are also supported by the BK21 project by the Korean Government.
KC thanks also the support of the Spanish MICINN°Øs
Consolider-Ingenio 2010 Programme under grant MultiDark
CSD2009-00064.

\section*{Appendix A: 1PI axino-gluino-gluon amplitude}

In this appendix, we provide an explicit computation of the 1PI
axino-gluino-gluon amplitude for a simple, but still general enough,
form of Wilsonian effective interaction of the axion superfield at a
generic cutoff scale $\Lambda<v_{PQ}$, which is given by
\begin{eqnarray}
\Delta_1{\cal L} &=& -\int d^2\theta ~ \frac{C_W}{32 \pi^2}\frac{A}{
 v_{PQ}}W^{a\alpha}W^a_\alpha + \textrm{h.c.}, \\
\Delta_2{\cal L} &=& \int d^4\theta ~
\frac{(A+A^\dagger)}{v_{PQ}}\left(\tilde y_1\Phi^\dagger_1\Phi_1 +
\tilde y_1^c\Phi^{c\dagger}_1\Phi_1^c+
\tilde y_2\Phi^\dagger_2\Phi_2 + \tilde y_2^c\Phi^{c\dagger}_2\Phi^c_2\right),\\
\Delta_3{\cal L} &=& -\int d^2\theta ~ \frac{A}{v_{PQ}}\Big[(\tilde
x_1 +\tilde x_1^c)M_1 \Phi_1 \Phi_1^c + (\tilde x_2+\tilde x_2^c)
M_2 \Phi_2 \Phi_2^c \Big] + \textrm{h.c.},
\end{eqnarray}
where $W^a_\alpha$ are the gluon superfields, $\Phi_i+\Phi_i^c$
($i=1,2$) form $3+\bar 3$ of $SU(3)_c$, and  $M_2\gg M_1$, but both
masses are well below the cutoff scale $\Lambda$. In this effective
theory, the PQ symmetry is realized as \bea A\rightarrow
A+iv_{PQ}\alpha, \quad \Phi_i \rightarrow e^{i\tilde x_i
\alpha}\Phi_i, \quad \Phi^c_i \rightarrow e^{i\tilde x_i^c
\alpha}\Phi^c_i,\eea and the corresponding PQ anomaly is given by
\bea C_{PQ}=C_W + \sum_i (\tilde x_i +\tilde x_i^c).\eea Under the
field redefinition \bea \Phi_i \rightarrow e^{z_i A/v_{PQ}} \Phi_i, \quad
\Phi^c_i \rightarrow e^{z_i^c A/v_{PQ}} \Phi_i^c,\eea the Wilsonian
couplings change as \bea \label{repa_appen} && C_W \rightarrow C_W +
\sum_i
(z_i+z_i^c), \nonumber \\
&& (\,\tilde y_i,\, \tilde y_i^c,\, \tilde x_i, \,\tilde x_i^c \,)
\rightarrow (\,\tilde y_i+z_i,\, \tilde y_i^c+z_i^c,\, \tilde
x_i-z_i, \,\tilde x_i^c-z_i^c\,).\eea
\begin{figure}
\begin{center}
\subfigure[\label{gga_0_a}]{\includegraphics[width=3.5cm]{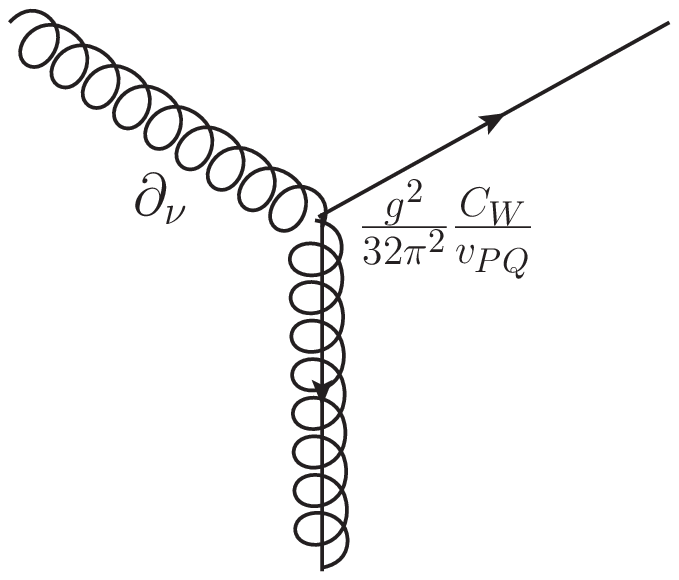}}
\quad
\subfigure[\label{gga_1_a}]{\includegraphics[width=3.5cm]{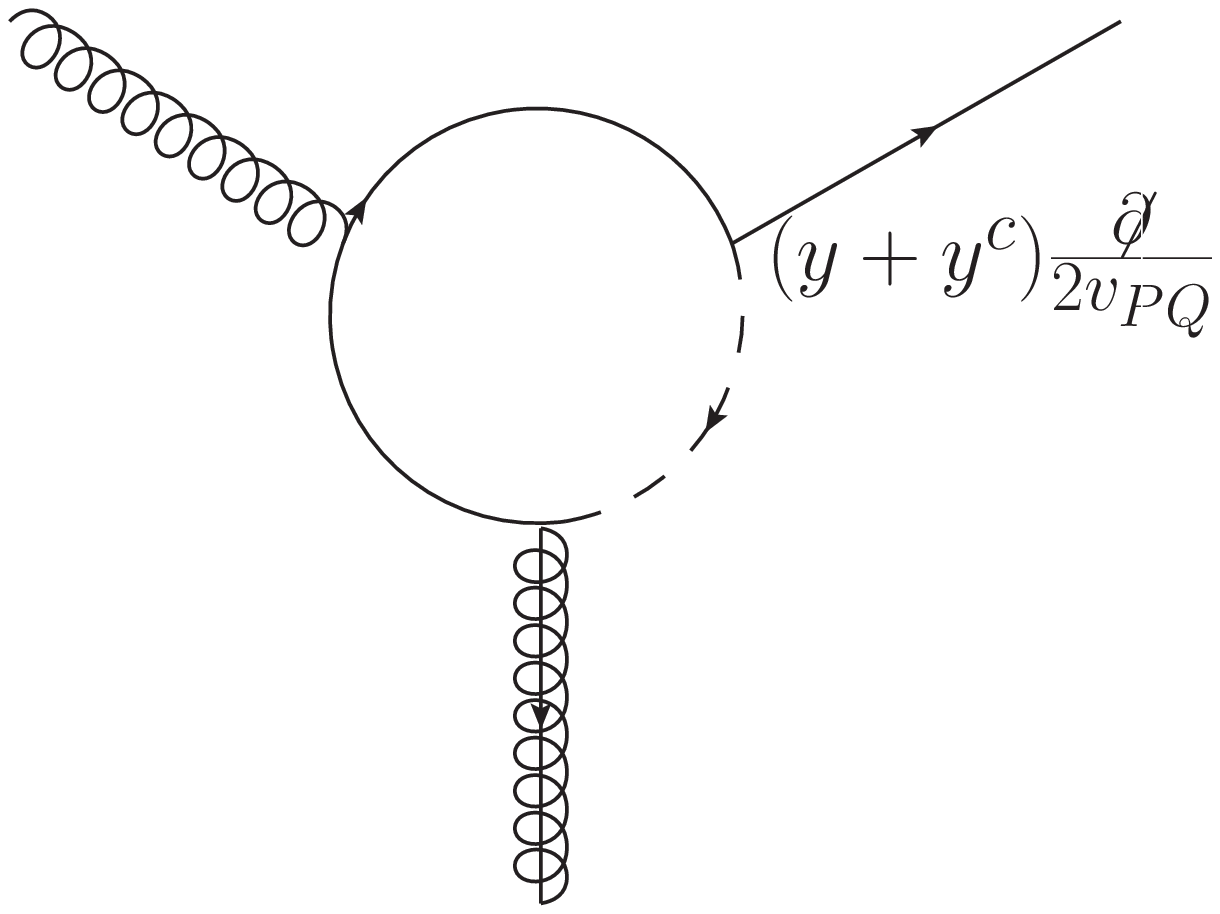}}
\quad
\subfigure[\label{gga_2_a}]{\includegraphics[width=3.5cm]{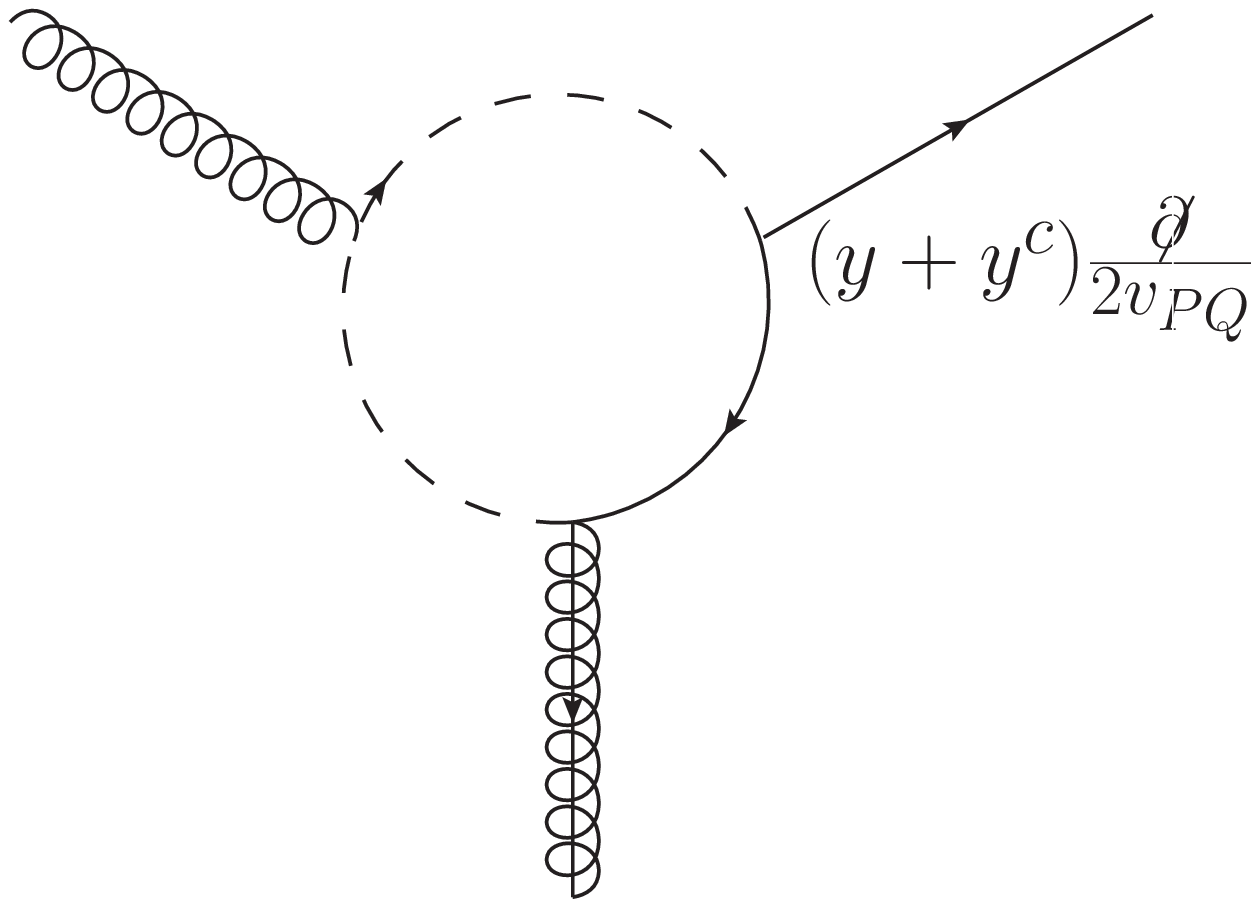}}
\\
\subfigure[\label{gga_3_a}]{\includegraphics[width=3.5cm]{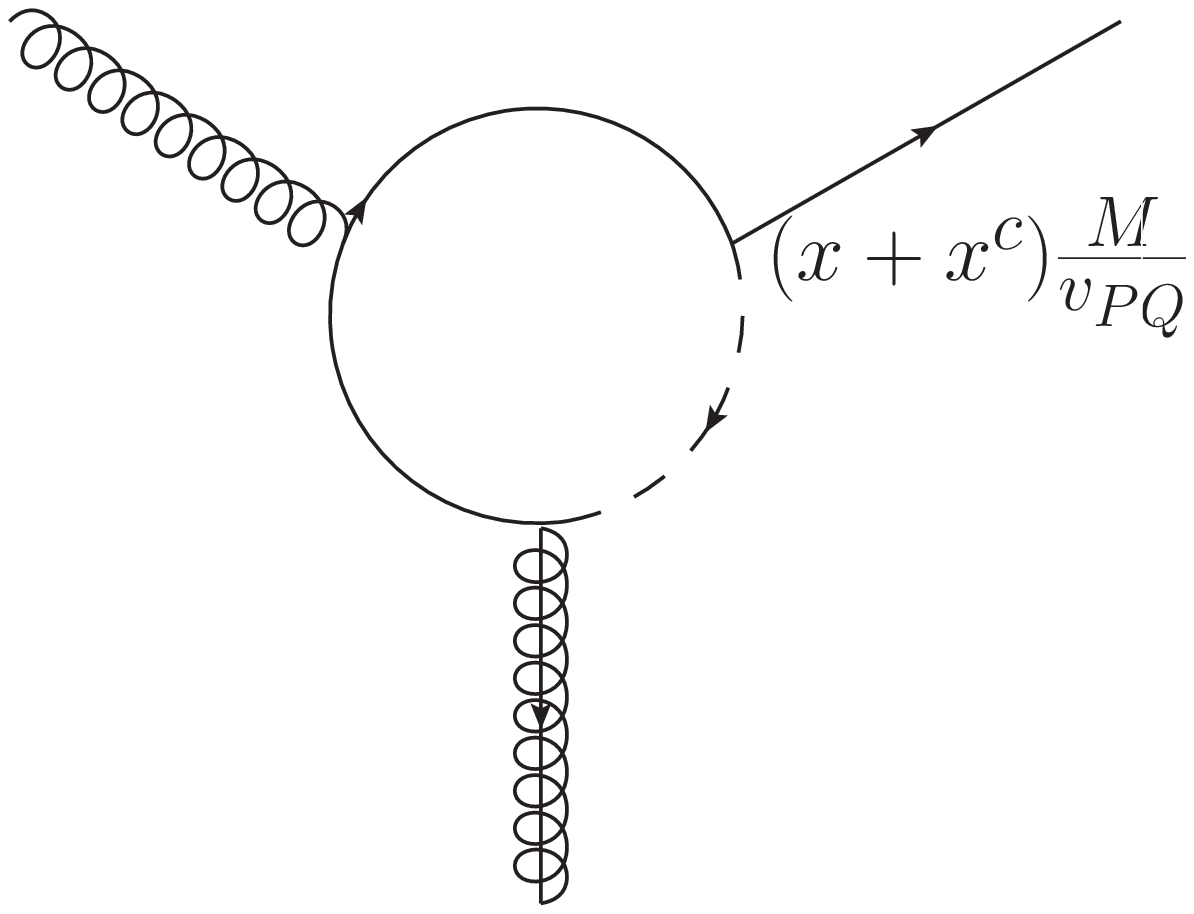}}
\quad
\subfigure[\label{gga_4_a}]{\includegraphics[width=3.5cm]{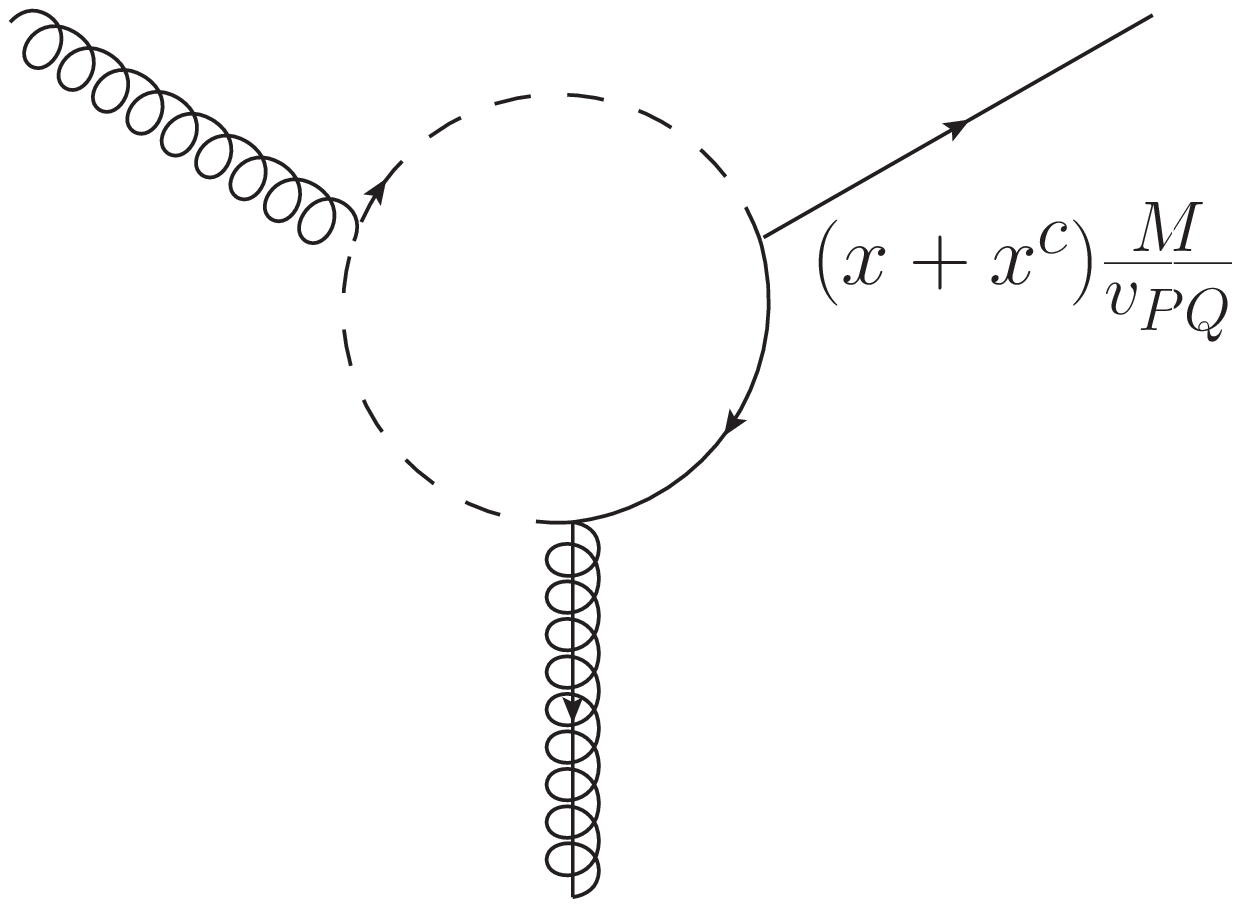}}
\caption{Axino-gluon-gluino amplitudes from the general effective interactions \label{gga_a}}
\end{center}
\end{figure}
 It is straightforward to compute the 1PI
axino-gluino-gluon amplitude from the above effective interactions,
yielding (see Fig. \ref{gga_a})\bea {\cal A}_{1PI}(k,q,p) =
-\frac{g^2}{16\pi^2 \sqrt{2}v_{PQ}} \delta^4(k+q+p) \widetilde
C_{1PI}(k,q,p)\bar{u}(k)\sigma_{\mu\nu} \gamma_5 v(q)\epsilon^\mu
p^\nu, \eea where $k, q$ and $p$ are the 4-momenta of the axino,
gluino and gluon, respectively, and\bea \widetilde C_{1PI} = C_W -\sum_i
(\tilde y_i+\tilde y_i^c)+\sum_i (\tilde y_i+\tilde y_i^c+\tilde
x_i+\tilde x_i^c)F(p,q;M_i)\eea for
\begin{equation}
F(p,q;M) \equiv \int^1_0 dx \int^{1-x}_0 dy
\frac{2M^2}{M^2-[p^2x(1-x)+q^2y(1-y) +2(p \cdot q)xy]}.
\end{equation}
Note that $\widetilde C_{1PI}$ is invariant under the reparametrization
(\ref{repa_appen}) as it should be. For the axino production by
gluon supermultiplet, the relevant kinematic situation is that axino
and gluino (or gluon) are on mass-shell, i.e. $k^2=q^2=0$ in the
limit to ignore SUSY breaking effects, while gluon (or gluino) is
off the mass-shell with $p^2=(k+q)^2<0$.
 In such kinematic region,
$F(p,q,M)$ is given by \bea F(k^2=q^2=0, \, p^2\neq 0)
=\frac{M^2}{|p^2|}\left[\log{\left(\frac{1+\sqrt{1-4M^2/|p^2|}}{-1+\sqrt{1-4M^2/|p^2|}}\right)}\right]^2,\eea
which can be approximated as \bea F & \simeq &
\frac{M^2}{|p^2|}\ln^2 \left(\frac{M^2}{|p^2|}\right)
\quad \textrm{for} \quad |p^2| \gg M^2 \\
F & \simeq & 1-\frac{1}{12}\frac{|p^2|}{M^2} \quad \textrm{for}
\quad |p^2|\ll M^2. \eea We then have \bea \widetilde
C_{1PI}(k^2=q^2=0,\, p^2\neq 0)= C_{1PI}+{\cal
O}\left(\frac{p^2}{M_{\rm heavy}^2}\right) +{\cal
O}\left(\frac{M_{\rm light}^2}{p^2}\ln^2\left(\frac{p^2}{M_{\rm
light}^2}\right)\right)\eea with
\begin{eqnarray}
\label{c1pi} && C_{1PI}(M_2^2<p^2<\Lambda^2) = C_W-(\tilde
y_1+\tilde y_1^c+ \tilde y_2+\tilde y_2^c),
 \nonumber \\
&& C_{1PI}(M_1^2<p^2<M_2^2) = C_W+(\tilde x_2+\tilde x_2^c-\tilde
y_1-\tilde y_1^c),
\nonumber
\\
&& C_{1PI}(\, p^2<M_1^2\,) = C_W+(\tilde x_1+\tilde x_1^c+ \tilde
x_2+\tilde x_2^c).
\end{eqnarray}
If $\Phi_2+\Phi_2^c$ corresponds to the heaviest PQ-charged and
gauge-charged matter superfield in the underlying model, we would
have the boundary condition \bea C_{1PI}(M_2^2<p^2<\Lambda^2)=0,\eea
which yields the following expression of  $C_{1PI}$ at generic
momentum scale below $v_{PQ}$: \bea C_{1PI}(p) = \sum_{M_i^2>|p^2|}
(\tilde x_i+\tilde x_i^c+\tilde y_i+\tilde y_i^c).\eea

Finally, we consider the expression of $\widetilde C_{1PI}$ in another
kinematic situation  when one of the 4-momenta $p,q,k$ is vanishing:
\bea F(k=0, \, p^2=q^2\neq 0)=
\frac{2M^2/|p^2|}{\sqrt{1+4M^2/|p^2|}}\ln{\left(\frac{1+\sqrt{1+4M^2/|p^2|}}{-1+\sqrt{1+4M^2/|p^2|}}\right)}
\eea which has the limiting behavior \bea F&\simeq&
-\frac{2M^2}{|p^2|}\ln\left(\frac{M^2}{|p^2|}\right)\quad \mbox{for}
\quad |p^2| \gg M^2  \nonumber \\
F & \simeq & 1-\frac{1}{6}\frac{|p^2|}{M^2} \quad \mbox{for}\quad
|p^2| \ll M^2,
\end{eqnarray}
and therefore
 \bea \widetilde C_{1PI}(k=0,\, p^2=q^2\neq 0) = C_{1PI}+{\cal
O}\left(\frac{p^2}{M_{\rm heavy}^2}\right) +{\cal
O}\left(\frac{M_{\rm light}^2}{p^2}\ln\left(\frac{p^2}{M_{\rm
light}^2}\right)\right),\eea where $C_{1PI}^a$ are given as
(\ref{c1pi}).

\section*{Appendix B: RG running and threshold corrections}

In this appendix, we discuss  the RG running of the Wilsonian  and
1PI amplitudes of the axion superfield, using the method of analytic
continuation into $N=1$ superspace.
For this, let us consider a generic Wilsonian effective lagrangian
\bea \label{eff_appen} {\cal L}_W(\Lambda) &=& \int d^4\theta\,
Z_n(\Lambda;A+A^\dagger) \Phi_n^\dagger \Phi_n \\
&&+ \int d^2\theta
\left[\,\frac{1}{4}f^{\rm eff}_a(\Lambda;A) W^{a\alpha}W^a_\alpha \right.+ \left. \frac{1}{2}e^{-(\tilde x_m+\tilde
x_n)A/v_{PQ}}M_{mn}\Phi_m\Phi_n\right.\nonumber \\
&&+\left.\frac{1}{6}e^{-(\tilde x_m+\tilde
x_n+ \tilde x_p)A/v_{PQ}}\lambda_{mnp}\Phi_m\Phi_n\Phi_p +{\rm h.c.}\right], \eea for
which $C_W$ and $\tilde y_n$ can be defined as \bea C_W(\Lambda)
=-8\pi^2 v_{PQ}\left.\frac{\partial f^{\rm eff}_a}{\partial
A}\right|_{A=0}, \quad \tilde y_n(\Lambda) =v_{PQ}\left.\frac{d \ln
Z_n}{\partial A}\right|_{A=0}. \eea We then have \bea \frac{d \tilde
x_n}{d\ln\Lambda} &=& 0, \nonumber \\
 \frac{d
C_W}{d\ln\Lambda}&=& -8\pi^2 v_{PQ}\left.\frac{\partial}{\partial
A}\left(\frac{d f^{\rm eff}_a}{d\ln\Lambda}\right) \right|_{A=0} =
0,\eea while $\tilde y_n$ can have a nontrivial RG running as \bea
\frac{d \tilde y_n}{d\ln\Lambda} &=&
v_{PQ}\left.\frac{\partial}{\partial A}\left(\frac{d \ln
Z_n}{d\ln\Lambda}\right) \right|_{A=0},
\eea
where
 we have used that the RG running
of the holomorphic gauge kinetic function $f^{\rm eff}_a$ is
exhausted at one-loop, and therefore $df^{\rm eff}_a/d\ln \Lambda =
b_a/16\pi^2$ is an $A$-independent constant.

In the appendix A, we have seen that the 1PI axino-gaugino-gauge
boson amplitude $\widetilde C^a_{1PI}$ in the kinematic regime with
$k=0$ and $M_1^2<p^2<M_2^2$ is given by \bea \widetilde C_{1PI}=
C_{1PI}+{\cal O}\left(\frac{p^2}{M_{2}^2}\right) +{\cal
O}\left(\frac{M_{1}^2}{p^2}\ln\left(\frac{p^2}{M_{1}^2}\right)\right),\eea
where $C_{1PI}$ is a constant in 1-loop approximation. Including
higher loops, $C_{1PI}$ can have a logarithmic $p$-dependence, which
can be determined by the 1PI RG equation. To examine this, let us
consider the 1PI effective lagrangian at a momentum scale
$p<\Lambda$, including \bea {\cal L}_{1PI}=\int d^4\theta \,
\left[\,{\cal Z}_n(p, A+A^\dagger)\Phi_n^\dagger \Phi_n
+\frac{1}{16}{\cal F}_a(p,
A+A^\dagger)\left(W^{a\alpha}\frac{D^2}{p^2}W^a_\alpha +{\rm
h.c.}\right)\,\right],\eea where ${\cal Z}_n$ is the 1PI wavefunction
coefficient chosen to satisfy the matching condition \bea {\cal
Z}_n(p^2=\Lambda^2)=Z_n(\Lambda),\eea where $Z_n(\Lambda)$ is the
Wilsonian wavefunction coefficient at the cutoff scale $\Lambda$. We
then  have \bea C_{1PI}=-8\pi^2 v_{PQ}\left.\frac{\partial {\cal
F}_a}{\partial A}\right|_{A=0}.\eea One can introduce also the
canonical 1PI Yukawa couplings \bea \Lambda_{mnp} \equiv
\frac{e^{-(\tilde x_m+\tilde x_n+ \tilde
x_p)A/v_{PQ}}\lambda_{mnp}}{\sqrt{{\cal Z}_m{\cal Z}_n{\cal
Z}_p}}\eea which obeys \bea v_{PQ}\left.\frac{\partial
|\Lambda_{mnp}|^2}{\partial A}\right|_{A=0}=-(\tilde x_m+\tilde x_n
+\tilde x_p +\tilde y_m(p) +\tilde y_n(p) + \tilde
y_p(p))|\Lambda_{mnp}|^2,\eea where \bea \tilde y_n (p) =
v_{PQ}\left.\frac{\partial \ln {\cal Z}_n(p)}{\partial
A}\right|_{A=0}.\eea

The 1PI gauge kinetic function  obeys the
Novikov-Shifman-Vainshtein-Zakhatov (NSVZ) RG equation \bea
\frac{d{\cal F}_a}{d\ln p^2}  \equiv
\beta_a=\frac{1}{16\pi^2}\frac{1}{1-{\rm Tr}(T_a^2(G))/8\pi^2{\cal
F}_a}\left(b_a +\sum_n {\rm Tr}(T_a^2(\Phi_n))\gamma_n \right),\eea
where \bea \gamma_n(p) \equiv \frac{d\ln {\cal Z}_n}{d\ln p}.\eea
 Both $\beta_a$ and $\gamma_n$ can be expressed as a function of
 the 1PI gauge coupling $g_a^2=1/{\cal F}_a$ and the 1PI Yukawa coupling
 $|\Lambda_{mnp}|^2$.
We then find the 1PI RG equations for $C^a_{1PI}$ and $\tilde y_n$
are given by
  \bea \label{rg1}
  \frac{d C_{1PI}}{d\ln p} &=& -8\pi^2 v_{PQ}\left.\frac{\partial
\beta_a}{\partial A}\right|_{A=0}\, =\,
-8\pi^2v_{PQ}\left.\left(\frac{\partial \beta_a}{\partial {\cal
F}_a}\frac{\partial {\cal F}_a}{\partial A}+\frac{\partial
\beta_a}{\partial |\Lambda_{mnp}|^2}\frac{\partial
|\Lambda_{mnp}|^2}{\partial A}\right)\right|_{A=0}\nonumber \\
&=& C_{1PI}^a\frac{\partial \beta_a}{\partial {\cal
F}_a}+8\pi^2(\tilde x_m+\tilde x_n +\tilde x_p +\tilde y_m +\tilde
y_n + \tilde y_p)|\Lambda_{mnp}|^2\frac{\partial
\beta^a_{1PI}}{\partial |\Lambda_{mnp}|^2}, \\
\label{rg2} \frac{d \tilde y_n}{d\ln p} &=&
v_{PQ}\left.\frac{\partial \gamma_n}{\partial A}\right|_{A=0}
=v_{PQ}\left.\left(\frac{\partial \gamma_n}{\partial {\cal
F}_a}\frac{\partial {\cal F}_a}{\partial A}+\frac{\partial
\gamma_n}{\partial |\Lambda_{mnp}|^2}\frac{\partial
|\Lambda_{mnp}|^2}{\partial
A}\right)\right|_{A=0}\nonumber \\
&=&\frac{C_{1PI}^ag_a^4}{8\pi^2}\frac{\partial \gamma_n}{\partial g_a^2} -\,(\tilde x_m+\tilde x_n +\tilde x_p +\tilde y_m +\tilde y_n +
\tilde y_p)|\Lambda_{mnp}|^2\frac{\partial \gamma_n}{\partial
|\Lambda_{mnp}|^2}. \eea

In section 2, we noticed that the Yukawa couplings of the model are
constrained by the PQ selection rule \bea (\tilde x_m +\tilde x_n
+\tilde x_p +\tilde y_m +\tilde y_n +\tilde y_p)\lambda_{mnp}= {\cal
O}\left(\frac{M_{\Phi}^2}{v_{PQ}^2}\right)+{\cal
O}\left(\frac{v_{PQ}}{M_*}\right).\eea With the boundary value
$C_{1PI}^a$ at $p\simeq v_{PQ}$, which is estimated as (\ref{boun}),
 the RG equations
(\ref{rg1}) and (\ref{rg2}) assure that \bea
C_{1PI}^a(M_\Phi<p<v_{PQ})
 = {\cal O}\left(\frac{M_\Phi^2}{v_{PQ}^2}\right)+ {\cal
O}\left(\frac{v_{PQ}^2}{M_*^2}\right),\eea and therefore the result
(\ref{1pi_zero}) remains valid even after higher loop effects are
included.

\section*{Appendix C: Thermal axino production by matter multiplet}
In this appendix, we present a calculation of the relic axino
density when the axinos are produced dominantly by the matter
multiplet $\Phi$ at $T \gg M_\Phi$.
The Boltzmann equation
of axino production is given by
\begin{equation}
\frac{d n_{\tilde{a}}}{dt}+3Hn_{\tilde{a}}=
\langle\sigma_{(I+J\to K+\tilde{a})}v\rangle (n_I n_J-n_K
n_{\tilde{a}}),
\end{equation}
where $n_I$ is the number density of the $I$-th particle species. At
high temperature of our interest, all particles except axinos are
in thermal equilibrium, so we can write $n_I=n_I^{\text{eq}}$, where
$n_I^{\text{eq}}$ is the equilibrium number density of particle $I$.
From the relation of the detailed balance, we have
\begin{equation}
\langle\sigma_{(I+J\to K+\tilde{a})}v\rangle n_I^{\text{eq}}n_J^{\text{eq}}=
\langle\sigma_{(I+J\to K+\tilde{a})}v\rangle n_K^{\text{eq}}n_{\tilde{a}}^{\text{eq}}.\label{eq:detail}
\end{equation}
Defining $Y_{\tilde{a}}=n_{\tilde{a}}/s$, the Boltzmann equation becomes
\begin{equation}
\frac{ dY_{\tilde{a}}}{dt}=
s\langle\sigma_{(I+J\to K+\tilde{a})}v\rangle
Y_K^{\text{eq}}(Y_{\tilde{a}}^{\text{eq}}-Y_{\tilde{a}}).
\end{equation}
For $\Delta_{\tilde{a}}\equiv
Y_{\tilde{a}}^{\text{eq}}-Y_{\tilde{a}}$, we have
\begin{equation}
\frac{d \Delta_{\tilde{a}}}{\Delta_{\tilde{a}}}
=-s\langle\sigma_{(I+J\to K+\tilde{a})}v\rangle Y_K^{\text{eq}}
dt,
\end{equation}
which yields
\begin{equation}
\Delta_{\tilde{a}}(T_0)=\Delta_{\tilde{a}}(T_R)
\exp\biggl[-\int^{T_R}_{T_0}
\frac{s\langle\sigma_{(I+J\to K+\tilde{a})}v\rangle Y_K^{\text{eq}}}{H(T)T}dT
\biggr].
\end{equation}
As long as axinos are relativistic in the temperature of interest,
$Y_{\tilde{a}}^{\text{eq}}$ is independent of temperature. Then
 we can use
\begin{equation}
\begin{split}
\Delta_{\tilde{a}}(T_0)=&\Delta_{\tilde{a}}(T_R)
\exp\biggl[-\frac{1}{Y_{\tilde{a}}^{\text{eq}}}\int^{T_R}_{T_0}
\frac{s\langle\sigma_{(I+J\to K+\tilde{a})}v\rangle Y_K^{\text{eq}}Y_{\tilde{a}}^{\text{eq}}}{H(T)T}dT
\biggr]\\
=&\Delta_{\tilde{a}}(T_R)
\exp\biggl[-\frac{1}{Y_{\tilde{a}}^{\text{eq}}}\int^{T_R}_{T_0}
\frac{\langle\sigma_{(I+J\to K+\tilde{a})}v\rangle n_K^{\text{eq}}n_{\tilde{a}}^{\text{eq}}}{s(T)H(T)T}dT\biggr]\\
=&\Delta_{\tilde{a}}(T_R)
\exp\biggl[-\frac{1}{Y_{\tilde{a}}^{\text{eq}}}\int^{T_R}_{T_0}
\frac{\langle\sigma_{(I+J\to K+\tilde{a})}v\rangle
n_I^{\text{eq}}n_J^{\text{eq}}}{s(T)H(T)T}dT\biggr],
\end{split}
\end{equation}
where the relation (\ref{eq:detail}) is used for the last identity.
Under the initial condition
$\Delta_{\tilde{a}}(T_R)=Y_{\tilde{a}}^{\text{eq}}-Y_{\tilde{a}}(T_R)=Y_{\tilde{a}}^{\text{eq}}$,
the above result gives \bea
Y_{\tilde{a}}(T_0)&=&Y_{\tilde{a}}^{\text{eq}} \biggl\{
1-\exp\biggl[-\frac{1}{Y_{\tilde{a}}^{\text{eq}}}\int^{T_R}_{T_0}
\frac{\langle\sigma_{(I+J\to K+\tilde{a})}v\rangle
n_I^{\text{eq}}n_J^{\text{eq}}}{s(T)H(T)T}dT\biggr] \biggl\}
\nonumber \\
&\simeq  &
\int^{T_R}_{T_0} \frac{\langle\sigma_{(I+J\to
K+\tilde{a})}v\rangle
n_I^{\text{eq}}n_J^{\text{eq}}}{s(T)H(T)T}dT \nonumber \\
&=& \frac{\bar{g}M_{Pl}}{16\pi^4}\int^{\infty}_{t_R}dt
~t^3K_1(t) \int^{tT_R}_{(m_I+m_J)}d(\sqrt{s})\sigma(s)
\biggl[\frac{(s-m_I^2-m_J^2)^2-4m_I^2m_J^2}{s^2}\biggr],
\nonumber \eea where $\bar{g}=135\sqrt{10}/(2\pi^3g_*^{3/2})$,
$M_{Pl}=2.4\times10^{18}$ GeV, $t_R=(m_I+m_J)/T_R$, and $K_1(t)$ is
the modified Bessel function  \cite{choi}. The cross section
$\sigma(s)$ can be obtained by
\begin{equation}
\biggl(\frac{d\sigma}{d\Omega}\biggr)
=\frac{1}{2E_{p_1}2E_{p_2}|v_{p_1}-v_{p_2}|}
\frac{|\bf{k}_1|}{(2\pi)^24E_{\text{cm}}}
\big|{\cal M}(p_1p_2\to k_1k_2)\big|^2
\end{equation}
with the amplitudes listed in  Table \ref{axino:proc}. Then, for
$T_R\gg M_\Phi\gg M_{\tilde{g}}$, we find \bea
Y_{\tilde{a}}^{\Delta_3{\cal L}}(T_0) &\simeq
&\frac{\bar{g}M_{Pl}}{16\pi^4}\biggl(\frac{8g^2
(x_\Phi+x_{\Phi^c})^2
M_\Phi^2|T_{ij}^a|^2}{v_{PQ}^2}\biggr)\frac{1}{M_\Phi}\nonumber\\
&&\times\biggl(
\frac{7}{60}+\frac{3}{140}+0.14+\frac{11}{60}+0.12+0.28+0.17\biggr)\nonumber \\
&\simeq
&(x_\Phi+x_{\Phi^c})^2\biggl(\frac{\bar{g}M_{Pl}}{2\pi^4}\biggr)\biggl(\frac{g^2M_\Phi}{v_{PQ}^2}\biggr)\biggl(\frac{N^2-1}{2}\biggr),
\label{axino:numd} \eea where  $|T_{ij}^a|^2 = {\rm Tr}(T^a T^a) =
(N^2-1)/2$ for the $SU(N)$ gauge group, and the numbers in the first
line come from the integrations for the processes C, D, E, G, I, H
and J, respectively, in Table \ref{axino:proc}. From the above
result, the relic axino mass density can be determined   as \bea
\Omega_{\tilde{a}}h^2 = \rho_{\tilde{a}}h^2/\rho_c =
m_{\tilde{a}}n_{\tilde{a}}h^2/\rho_c =
m_{\tilde{a}}Y_{\tilde{a}}s(T_0)h^2/\rho_c \simeq 2.8 \times
10^5
\biggl(\frac{m_{\tilde{a}}}{\textrm{MeV}}\biggr)Y_{\tilde{a}},
\eea where we have used  $\rho_c/[s(T_0)h^2] = 3.6 \times 10^{-9}$
GeV.

\end{document}